\documentclass[aps,prb,twocolumn,amsmath,amssymb,superscriptaddress,citeautoscript,longbibliography,floatfix]{revtex4-2}

\usepackage{natbib}
\usepackage{color}
\usepackage{graphicx}
\usepackage{hyperref}

\begin{document}

\title{\boldmath The planar triangular $S=3/2$ magnet AgCrSe$_2$: magnetic frustration, short range correlations, and field tuned anisotropic cycloidal magnetic order}


\author{M.~Baenitz}\email[Corresponding author: \vspace{5pt}]{Michael.Baenitz@cpfs.mpg.de}
\affiliation{Max Planck Institute for Chemical Physics of Solids, D-01187 Dresden, Germany}

\author{M.~M.~Piva}
\affiliation{Max Planck Institute for Chemical Physics of Solids, D-01187 Dresden, Germany}

\author{S.~Luther}
\affiliation{Hochfeld-Magnetlabor Dresden (HLD-EMFL) and W\"{u}rzburg-Dresden Cluster of Excellence ct.qmat, Helmholtz-Zentrum Dresden-Rossendorf, 01328 Dresden, Germany}
\affiliation{Institut für Festk\"{o}rper- und Materialphysik, TU Dresden, 01062 Dresden Germany}

\author{J.~Sichelschmidt}
\affiliation{Max Planck Institute for Chemical Physics of Solids, D-01187 Dresden, Germany}

\author{K.~M. Ranjith}
\affiliation{Max Planck Institute for Chemical Physics of Solids, D-01187 Dresden, Germany}

\author{H.~Dawczak-D\c{e}bicki}
\affiliation{Max Planck Institute for Chemical Physics of Solids, D-01187 Dresden, Germany}

\author{M.~O. Ajeesh}
\affiliation{Max Planck Institute for Chemical Physics of Solids, D-01187 Dresden, Germany}

\author{S.-J.~Kim}
\affiliation{Max Planck Institute for Chemical Physics of Solids, D-01187 Dresden, Germany}

\author{G.~ Siemann}
\affiliation{School of Physics and Astronomy, University of St. Andrews, St. Andrews KY16 9SS, United Kingdom}

\author{C.~Bigi}
\affiliation{School of Physics and Astronomy, University of St. Andrews, St. Andrews KY16 9SS, United Kingdom}

\author{P.~Manuel}
\affiliation{ISIS Neutron and Muon Source, Rutherford Appelton Laboratory, Chilton, Didcot OX11 OQX, United Kingdom}

\author{D.~Khalyavin}
\affiliation{ISIS Neutron and Muon Source, Rutherford Appelton Laboratory, Chilton, Didcot OX11 OQX, United Kingdom}

\author{D.~A.~Sokolov}
\affiliation{Max Planck Institute for Chemical Physics of Solids, D-01187 Dresden, Germany}

\author{P.~Mokhtari}
\affiliation{Max Planck Institute for Chemical Physics of Solids, D-01187 Dresden, Germany}

\author{H.~Zhang}
\affiliation{Max Planck Institute for Chemical Physics of Solids, D-01187 Dresden, Germany}

\author{H.~Yasuoka}
\affiliation{Max Planck Institute for Chemical Physics of Solids, D-01187 Dresden, Germany}

\author{P.~D.~C.~ King}
\affiliation{School of Physics and Astronomy, University of St. Andrews, St. Andrews KY16 9SS, United Kingdom}

\author{G.~Vinai}
\author{V.~Polewczyk}
\author{P.~Torelli}
\affiliation{Istituto Officina dei Materiali (IOM)-CNR, Laboratorio TASC, Area Science Park, S.S. 14 km 163.5, Trieste I-34149, Italy}

\author{J.~Wosnitza}
\affiliation{Hochfeld-Magnetlabor Dresden (HLD-EMFL) and W\"{u}rzburg-Dresden Cluster of Excellence ct.qmat, Helmholtz-Zentrum Dresden-Rossendorf, 01328 Dresden, Germany}
\affiliation{Institut für Festk\"{o}rper- und Materialphysik, TU Dresden, 01062 Dresden Germany}

\author{U.~Burkhardt}
\affiliation{Max Planck Institute for Chemical Physics of Solids, D-01187 Dresden, Germany}

\author{B.~Schmidt}
\affiliation{Max Planck Institute for Chemical Physics of Solids, D-01187 Dresden, Germany}

\author{H.~Rosner}
\affiliation{Max Planck Institute for Chemical Physics of Solids, D-01187 Dresden, Germany}

\author{S.~Wirth}
\affiliation{Max Planck Institute for Chemical Physics of Solids, D-01187 Dresden, Germany}

\author{H.~K\"{u}hne}
\affiliation{Hochfeld-Magnetlabor Dresden (HLD-EMFL) and W\"{u}rzburg-Dresden Cluster of Excellence ct.qmat, Helmholtz-Zentrum Dresden-Rossendorf, 01328 Dresden, Germany}

\author{M.~Nicklas}
\affiliation{Max Planck Institute for Chemical Physics of Solids, D-01187 Dresden, Germany}

\author{M.~Schmidt}
\affiliation{Max Planck Institute for Chemical Physics of Solids, D-01187 Dresden, Germany}

\date{\today}

\begin{abstract}\noindent
Single crystals of the hexagonal triangular lattice compound AgCrSe$_{2}$ have been grown by chemical vapor transport. The crystals have been carefully characterized and studied by magnetic susceptibility, magnetization, specific heat and thermal expansion. In addition, we used Cr-electron spin resonance and neutron diffraction to probe the Cr-3$d^3$ -- magnetism microscopically. To obtain the electronic density of states, we employed  X-ray absorption  and resonant photoemission spectroscopy (resPES) in combination with density functional theory calculations.
Our studies evidence an anisotropic magnetic order below $T_N = 32$~K. Susceptibility data in small fields of about 1~T reveal an antiferromagnetic (AFM) type of order for $H \perp c$, whereas for $H \parallel c$ the data are reminiscent of a field-induced ferromagnetic (FM) structure. At low temperatures and for $H \perp c$, the field-dependent magnetization and AC susceptibility data evidence a metamagnetic transition at $H^+ = 5$~T, which is absent for $H \parallel c$. We assign this to a transition from a planar cycloidal spin structure at low fields to a planar fan-like arrangement above $H^+$. A fully ferromagnetically polarized state is obtained above the saturation field of $H_{\perp S} = 23.7$~T at 2~K with a magnetization of $M_s = 2.8$~$\mu_{\rm B}{\rm /Cr}$. For $H \parallel c$, $M(H)$ monotonously increases and saturates at the same $M_s$ value at $H_{\parallel S} = 25.1$~T at 4.2~K.
Above $T_N $, the magnetic susceptibility and specific heat indicate signatures of two dimensional (2D) frustration related to the presence of planar ferromagnetic and antiferromagnetic exchange interactions. We found a pronounced nearly isotropic maximum in both properties at about $T^* = 45$~K, which is a clear fingerprint of short-range correlations and emergent spin fluctuations.  Calculations based on a planar 2D Heisenberg model support our experimental findings and suggest a predominant FM exchange among nearest and AFM exchange among third-nearest neighbors. Only a minor contribution might be assigned to the antisymmetric Dzyaloshinskii-Moriya interaction possible related to  the non-centrosymmetric polar space group $R3m$. Due to these competing interactions, the magnetism in AgCrSe$_{2}$, in contrast to the oxygen based delafossites, can be tuned by relatively small, experimentally accessible, magnetic fields, allowing us to establish the complete anisotropic magnetic $H-T$ phase diagram in detail.

\end{abstract}

\maketitle

\section{INTRODUCTION}

Covalently bonded two dimensional (2D) structures are frequently found with the triangular motif realized by a simple planar hexagonal lattice, in particular, but also by more complex distinguished 2D planar geometries such as the honeycomb-lattice. Anisotropic metallic conductivity due to anisotropic hybridization of electronic states evolves through a bonding situation given by the crystal structure. The simplest examples are graphite and graphene crystallizing in a hexagonal and a honeycomb lattice, respectively. The latter exhibits very intriguing properties, such as high charge-carrier mobility due to the formation of Dirac nodes near the Fermi level in the electronic band structure \cite{Neto}. Other examples for inorganic 2D materials are hexagonal boron nitrides and transition-metal chalcogenides \cite{Butler}. The triangular lattice delafossite PdCoO$_{2}$ belongs to the latter material class, which is characterized by an unusually high in-plane charge-carrier mobility \cite{Moll,AP}. Some organic BETS- and ET-salts also exhibit a layered triangular structure. They become metallic and even superconducting under pressure. Here the molecular Mott insulator {$\kappa$-ET$_{2}$Ag$_{2}$(CN)$_{3}$} \cite{Shimizu} shows in a remarkable way a pressure-induced crossover from a spin liquid in the Mott insulator state to a correlated Fermi liquid in the metallic state. Planar magnetic triangular lattices with spin $1/2$ ions, such as Cu$^{2+}$, favor the evolution of a gapless spin liquid ground state, e.g.\ Ba$_{3}$CuSb$_{2}$O$_{9}$ \cite{Zhou1}, which was already predicted by Anderson 40 years ago \cite{Anderson}. Recently, 2D magnetic van der Waals magnets, such as Cr$_{2}$Ge$_{2}$Te$_{6}$, CrCl$_{3}$, or RuCl$_{3}$, gained a lot of attention in view of device applications
and fundamental research \cite{Burch}. Some of them are metallic with a rather high charge carrier mobility along the layers, e.g.\ GdTe$_{3}$ \cite{Lei}. Especially, if the structure lacks inversion symmetry a moderate spin-orbit interaction can promote the emergence of complex chiral magnetic textures, such as skyrmions \cite{Pfleiderer}.  Consequently, it is very appealing to study magnetic ions on a perfect triangular lattice, in a system with a strong 2D-like conductivity, which in addition lacks inversion symmetry.  Here, spin and charge degrees of freedom couple and, in the presence of a moderate spin-orbit interaction, spin polarized bands, anisotropic complex magnetic ordering, piezoelectricity, multiferroicity and an unconventional magnetotransport can be expected \cite{Wang,Fiebig,Manchon}.

All these preconditions are fulfilled in the system AgCrSe$_{2}$ which in an extended sense belongs to the class of delafossites. Only recently a study on AgCrSe$_{2}$ single crystals confirmed reasonably metallic behavior in the plane, perpendicular to the $c$ axis, with  $\rho_{\perp} \approx 3~{\rm m\Omega cm}$
and insulating behavior perpendicular to the plane, parallel to the $c$ axis,  with a resistivity ratio of about $\rho_{\parallel}/\rho_{\perp} \approx  100$ at 2~K \cite{Yano,Zhang}. This suggests that AgCrSe$_{2}$ is an anisotropic degenerated $p$-type semiconductor with a finite density of states (DOS) at the Fermi level, yielding a charge carrier density of about $7.6 \times 10^{19}~cm^{-3}$ at 300 K \cite{Yano}. Early experimental studies on polycrystalline samples combined with band-structure calculations already predicted in-plane ferromagnetism and spin polarized bands at the verge of metallicity \cite{Gautam}. In particular, it should be emphasized that, unlike most delafossites which crystallize in the rhombohedral  $\alpha$-NaFeO$_{2}$ structure, space group $R\bar3m$ \cite{AP,Cann}, AgCrSe$_{2}$ exhibits a hexagonal structure, with different Ag coordination, at room temperature. Moreover, AgCrSe$_{2}$ undergoes a structural transition at about 450~K from non-centrosymmetric polar $R3m$ symmetry at low temperatures to centrosymmetric $R\bar3m$ symmetry at high temperatures \cite{Lee,Li,Murphy}. Therefore, in strong contrast to most other delafossites, asymmetric exchange interactions between the Cr ions are possible in AgCrSe$_{2}$. This allows, in addition, to the Heisenberg superexchange and a possible weak Ruderman-Kittel-Kasuya-Yosida (RKKY) exchange,  a sizable Dzyaloshinskii-Moriya (DM) interaction between the magnetic ions \cite{DM}. This favors chiral magnetic phases in general and unconventional transport properties in particular. Prominent examples here are the B20 skyrmion compounds MnSi and FeGe \cite{Pfleiderer,MB}, the half Heusler compounds, such as Mn$_{3}$Sn, \cite{SA}, and the Cr-Nb-chalcogenide Cr$_{1/3}$NbS$_{2}$ \cite{Cao}. Furthermore, the spin-orbit interaction in combination with the polar space group supports the presence of a Rashba effect which is expected to have a strong impact on the electronic (magneto) transport.

Delafossite-type structures form as $A^{1+}R^{3+}X^{2-}_{2}$, where $A$ is an alkaline-metal (Li, Na, K, Rb, Cs) or a monovalent transition-metal ion (Pd, Pt, or Cu, Ag), $R$ is a trivalent transition metal or rare-earth ion which might be magnetic (Ti, V, Cr, Fe, Ce, or Yb) or nonmagnetic (Al, Ga, In, Tl, or Co, Rh), and $X$ stands for a chalcogen \cite{Cann}. So far, most of the research carried out was on oxygen-based delafossites, featuring a wide variety of topical phenomena, including superconductivity (Na$_{x}$CoO$_{2}$) \cite{Takada}, large thermoelectricity (Cu[Rh,Mg]O$_{2}$) \cite{Kur}, multiferroicity (CuFeO$_{2}$) \cite{Hara}, hydrodynamic 2D conductivity (PdCoO$_{2}$) \cite{Moll}, and spin-orbit driven frustration on a perfect triangular lattice (NaYbO$_{2}$) \cite{Ranjith}.

In that respect, it seems to be promising to move from the field of oxygen-based delafossites to the less studied non-oxygen delafossites with S, Se, or Te. The variation of the ionic radii of the mono- and trivalent ions mainly tunes the lattice parameters. The monovalent ion dominates the conductivity and has a strong impact on the density of states (DOS) at the Fermi level. The choice of the chalcogenide ion itself has a sizable impact on the band structure and the DOS. For example, the oxygen-based delafossite PdCoO$_{2}$ has a unique quasi-2D  conductivity and the DOS at the Fermi level has predominant Pd character and shows a minor chalcogenide $2p$-electron admixture. The same is true for its magnetic relative PdCrO$_{2}$ which is a rather unusual itinerant magnet with short-range correlations and a rather strong antiferromagnetic (AFM) exchange. Both systems are an exception among oxygen-based delafossites whereas others, such as CuCrO$_{2}$ and AgCrO$_{2}$, are found to be rather bad metals.  In contrast, in AgCrS$_{2}$ or AgCrSe$_{2}$ the spatially more extended $3(4)p$-states of S and Se promote the emergence of hybridized $p$-states near the Fermi level which finally leads to reasonable metallic conductivity \cite{Yano}. To some extent, this is also found in Cr chalcogenide spinel compounds \cite{Baltzer, Huang}, but the exchange here is three-dimensional which is in strong contrast to Cr- delafossites. The hybridization enhances correlation effects between spin and charge in general and together with the emergence of strong spin polarization, fosters unconventional magnetism and (magneto-) transport. 

As mentioned above, the non oxygen-based chromium delafossites are hardly investigated and most of the research was done on polycrystalline material. We have been able to grow large high quality single crystals using vapor transport techniques. The single crystals allow to study the temperature and field dependence of the  anisotropic magnetism and to map out the $H-T$ phase diagram of AgCrSe$_{2}$. Whereas the magnetism in chromium oxygen-based delafossites, e.g.\ CuCrO$_{2}$, is quite robust against applied magnetic fields and no saturation of the magnetization occurs up to 100~T \cite{Miyata}, we show that the magnetism of AgCrSe$_{2}$ can be easily tuned to saturation by magnetic fields in the range of a few tens of tesla and that the expected saturation magnetization for $S=3/2$ Cr$^{3+}$ close to $3$~$\mu_{\rm B}/{\rm Cr}$ is obtained. Furthermore, we combine our magnetization studies with heat capacity, thermal expansion, electron spin resonance, and neutron diffraction. The results are discussed in terms of a frustrated 2D triangular $S=3/2$ lattice with competing exchange interactions among nearest neighbors (ferromagnetic) and third-nearest neighbors (antiferromagnetic) within the plane. The parameters used in the presented exchange model provide a very good description of the experimentally determined saturation fields and Weiss temperature. The neutron diffraction data confirms the magnetic transition to a complex planar cycloidal magnetic order with a rather long period (about 164 \AA) in (1,1,0) direction \cite{Engelsman}. The temperature dependencies of specific heat, susceptibility and neutron Bragg-peak intensities together with the rather small residual entropy ($0.2{\rm R}\ln4$) point at the emergence of anisotropic, possibly chiral, fluctuations in the vicinity of magnetic order.

The manuscript is organized as follows. After the overview on triangular lattice delafossite type systems, we present the applied methods in Sec. II. In Sec. III, we describe and discuss the experimental results: crystal growth, electronic structure followed by the magnetic and thermodynamic bulk measurements (susceptibility, magnetization, specific heat and thermal expansion) up to the two local probes applied (ESR spectroscopy and neutron scattering). We then describe the exchange in a 2D Heisenberg model and present the $H-T$ phase diagram. Finally in Sec. IV, we summarize all conclusions and give an outlook on possible future experiments. 

\section{METHODS}\label{Sec:Methods}

Magnetization measurements were performed using magnetic property measurement systems (MPMS and MPMS3, Quantum Design) and the AC measurement system option of the physical property measurement system (PPMS, Quantum Design). The magnetization experiments in pulsed magnetic fields up to 50~T were performed at the Dresden High Magnetic Field Laboratory (HLD) with a compensated pickup-coil system in a pulse-ﬁeld magnetometer in a home-built $^4$He cryostat and a pulsed magnet with an inner bore of 20~mm, powered by a 1.44~MJ capacitor bank \cite{Skourski}. Specific heat was recorded using the corresponding option of a PPMS. We investigated the electron-spin resonance (ESR) of single crystalline AgCrSe$_{2}$ using a standard continuous-wave ESR setup at X-band frequency (9.4~GHz). The temperature was varied between 3 and 300~K with a He-flow cryostat. ESR can be detected by the absorbed power $P$ of a transversal magnetic microwave field as a function of a static, external magnetic field $\mu_0H$. To improve the signal-to-noise ratio, we used a lock-in technique by modulating the static field, which yields the derivative of the resonance signal $dP/dH$. The measured ESR spectra were fitted with a Lorentzian function including the influence of the counter-rotating component of the linearly polarized microwave field \cite{rauch15a}. From the fit we obtained the linewidth $\Delta H$ and the resonance field $H_{res}$ which determines the ESR $g$-factor $g=h\nu/\mu_BH_{res}$. The ESR intensity $I_{\rm ESR}$ is a measure of the local static susceptibility of the probed ESR spin, i.e.\ in our case the local susceptibility of the Cr$^{3+}$ spins \cite{abragam70a}. We calculated $I_{ESR}\approx Amp\cdot \Delta H^{2}$ which approximates the integrated ESR absorption \cite{gruner10a}. The thermal expansion experiments were performed in a PPMS, equipped with a custom designed high-resolution dilatometer \cite{Kuechler}. Since temperature and magnetic field variations affects both, the sample length and the length of the Be-Cu dilatometer cell, a high purity copper plate (ADVENT Research Materials Ltd. 99.995\% Cu) of the same thickness as the AgCrSe$_2$ sample was used to calibrate the cell background. Neutron powder diffraction data were collected at the ISIS pulsed neutron and muon facility of the Rutherford Appleton Laboratory (UK) on the WISH diffractometer located at the second target station \cite{Chapon}. The sample was loaded into cylindrical 3~mm diameter vanadium cans and measured in the temperature range between $1.5$ and $100$~K (step size 5~K) using an Oxford Instruments cryostat. Rietveld refinements of the crystal and magnetic structures were performed using the FULLPROF program \cite{Rodriguez} against the data measured in detector banks at average $2\theta$ values of $27^{\circ}$, $58^{\circ}$, $90^{\circ}$, $122^{\circ}$, and $154^{\circ}$, each covering $32^{\circ}$ of the scattering plane.  X-ray absorption (XAS) and resonant photoemission spectroscopy (resPES) measurements were performed at the APE-HE beamline at Elletra sincrotrone, Italy. Samples were cleaved {\it in-situ} at a base pressure lower than $10^{-10}$ mbar. The measurements shown here were performed using photon energies in the range of $\sim\!569-596$~eV, with the sample held at a temperature of 100~K; very similar measurements were obtained at room temperature. XAS was performed in the total electron yield (TEY) mode. ResPES measurements were performed using a Scienta Omicron R3000 hemispherical electron energy analyzer. The measured Fermi edge of a gold reference sample was used for binding energy calibration for the resPES, as well as to reference the incident photon energy, assuming an analyser work function of 4.4~eV. The gold sample has been prepared via argon sputtering to observe a clean surface.  From the width of the measured Fermi edge of the gold reference sample, we estimate an experimental energy resolution of $\sim\!0.4$~eV.  DFT electronic structure calculations were performed using the full-potential FPLO code, version fplo18.00-52 (http://www.fplo.de) \cite{PhysRevB.59.1743,PhysRevB.60.14035}. The calculations were based on the localized density approximation (LDA) with Perdew-Wang-92 exchange-correlation functional \cite{PhysRevB.45.13244}. The correlation effect of the Cr-$3d$ shell was taken into account by the LDA+$U$ method in the atomic-limit flavour with $U$ in the range of $0.5 - 6.0$ eV. The spin orbit coupling effect was treated non-perturbatively by solving the full Kohn-Sham-Dirac equation \cite{ESCHRIG2004723}. The rhombohedral unit cell in the space group No.160 ($R3m$) was used, with the experimental lattice parameters of $a = b = 3.6798$~\r{A}, $c = 21.2250$~\r{A} \cite{VANDERLEE1989216}. The Brillouin zone was sampled with a $k$-mesh of $8000~k$-points ($20 \times 20 \times 20$ mesh, 1540 points in the irreducible wedge of the Brillouin zone). The experimentally-measured DOS is additionally broadened as compared to the calculations due to finite lifetime effects as well as the experimental energy resolution. To aid comparison, the calculated DOS has been convoluted with a Gaussian of full-width at half maximum equal to the experimental energy resolution of $0.4$\,eV.

\section{\label{sec:level1}RESULTS AND DISCUSSION}

\subsection{Single-crystal growth, crystal structure, and characterization} \label{Sec:characterization}

\begin{figure}[bt!]
\includegraphics[clip,width=\columnwidth]{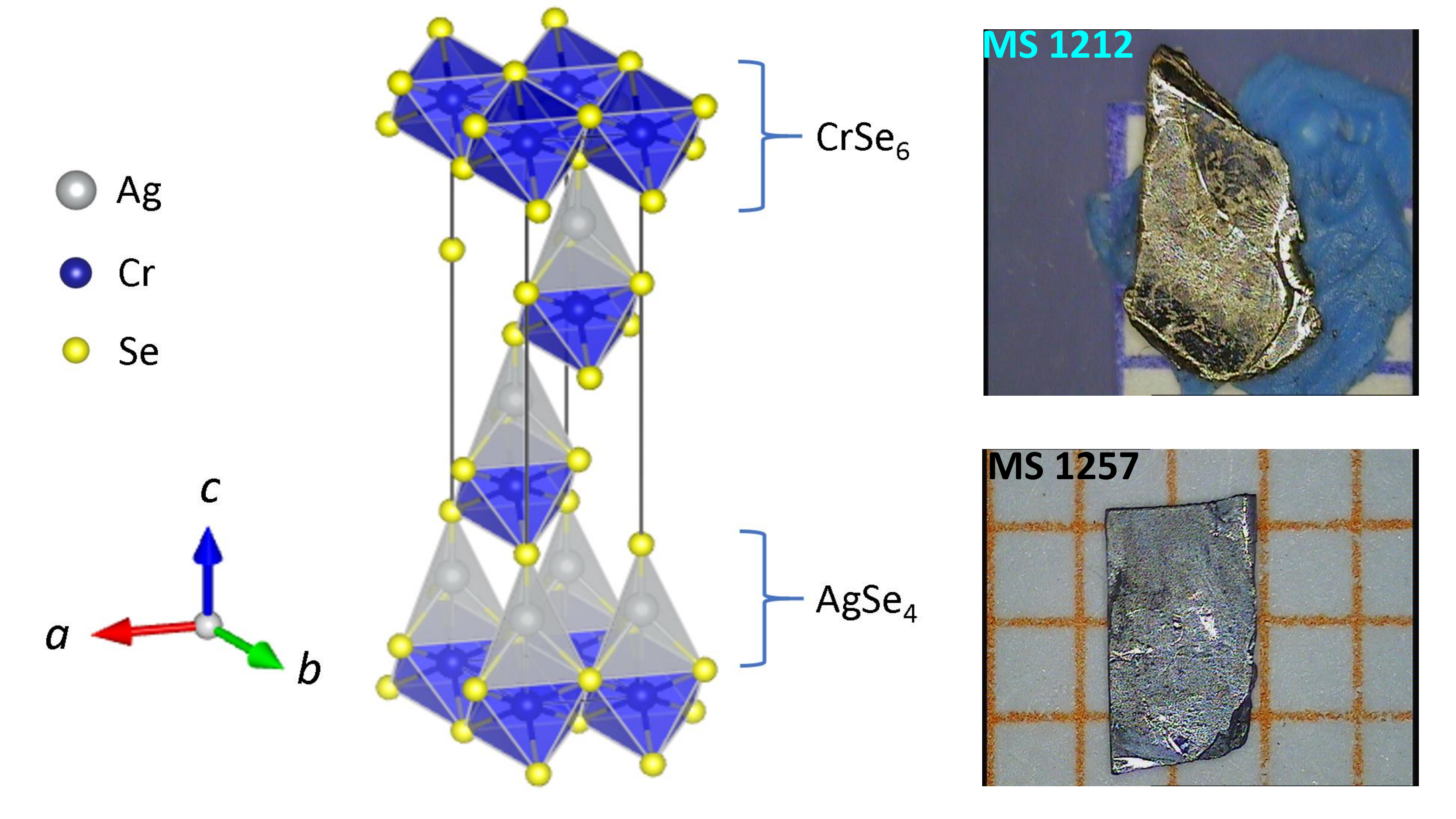}
\caption{Hexagonal lattice structure of AgCrSe$_{2}$ (left) and morphology of two crystals used in this study (right). The lateral extension corresponds to the ($a$,$b$) - plane and the direction perpendicular to it is the $c$ axis.}
\label{Crystal}
\end{figure}

We have grown high-quality AgCrSe$_{2}$ single crystals with lateral dimensions of a few millimeter via chemical transport reaction using chlorine as transport agent. In a first step polycrystalline AgCrSe$_{2}$ material has been synthesized by a direct reaction of the elements silver (powder 99.999\% Alfa Aesar), chromium (powder 99,99\% Alfa Aesar), and selenium (pieces 99.999\% Alfa Aesar, powdered) at $800^{\circ}$C in evacuated fused silica tubes within 14 days. Starting from this microcrystalline powder material, AgCrSe$_{2}$ crystallized by a chemical transport reaction in a temperature gradient from $900^{\circ}$C (source) to $800^{\circ}$C (sink), and a transport agent concentration of 3~mg/cm$^3$ chlorine. After 6 weeks, the experiment was stopped by quenching the ampoule in cold water. Selected crystals were characterized by energy-dispersive X-ray spectroscopy (EDXS), wavelength-dispersive X-ray spectroscopy (WDXS), X-ray powder diffraction, Laue X-ray diffraction, and differential scanning calorimetry (DSC). At 300~K, AgCrSe$_{2}$ crystallizes in a hexagonal structure (with $R3m$ crystal symmetry) with alternating AgSe$_{4}$ layers and CrSe$_{6}$ octahedral layers along the $c$ axis [see Fig.~\ref{Crystal} (left)] and lattice constants of $a=3.6824(3)$~\AA\ and $c=21.233(1)$~\AA, determined by powder X-ray diffraction. Figure~\ref{Crystal} (right panel) shows two crystals from different crystal growths. Magnetization measurements were performed on both crystals in great detail (see also the Supplemental Material \cite{SM}). High field magnetization measurements, electron spin resonance were performed on MS1212 and specific heat studies on MS1262 and MS1212.
Furthermore, the samples were investigated with respect to the structural phase transition between 350 and 450~K by differential-scanning calorimetry (DSC). Specific heat obtained by DSC was recorded on crystal MS1212 upon heating and cooling. The heating and cooling speed was 20~K/min. A hysteretic phase transition is clearly visible (see the Supplemental Material \cite{SM}).  The chemical composition of AgCrSe$_{2}$ crystals was studied locally by energy dispersive x-ray spectroscopy (EDXS) (details in the Supplemental Material \cite{SM}). Only rather small deviations from the nominal stoichiometry were found. Although the variations are small (and close to the experimental error), the trend is that the crystals all have a slightly lower silver content and an increased chromium content, which formally corresponds to Ag$_{0.92}$Cr$_{1.08}$Se$_{2}$. Especially the silver content seems to be important for the metallic conductivity as shown in a separate study \cite{Zhang}. A silver deficiency might lead to a higher metallic conductivity \cite{Tang} of the crystals. It seems that a rather slight silver deficiency is an inherent feature of our crystals. Initial resistivity and Hall measurements on our AgCrSe$_{2}$ crystals \cite{Zhang} and our ESR study (see Sec. \ref{SEC:ESR}) clearly indicate metallic conductivity, which is consistent with our EDX results and previous reports \cite{Tang}.

\subsection{Electronic structure} \label{SEC:electron}

\begin{figure}
	\centering
	\includegraphics[width=\columnwidth]{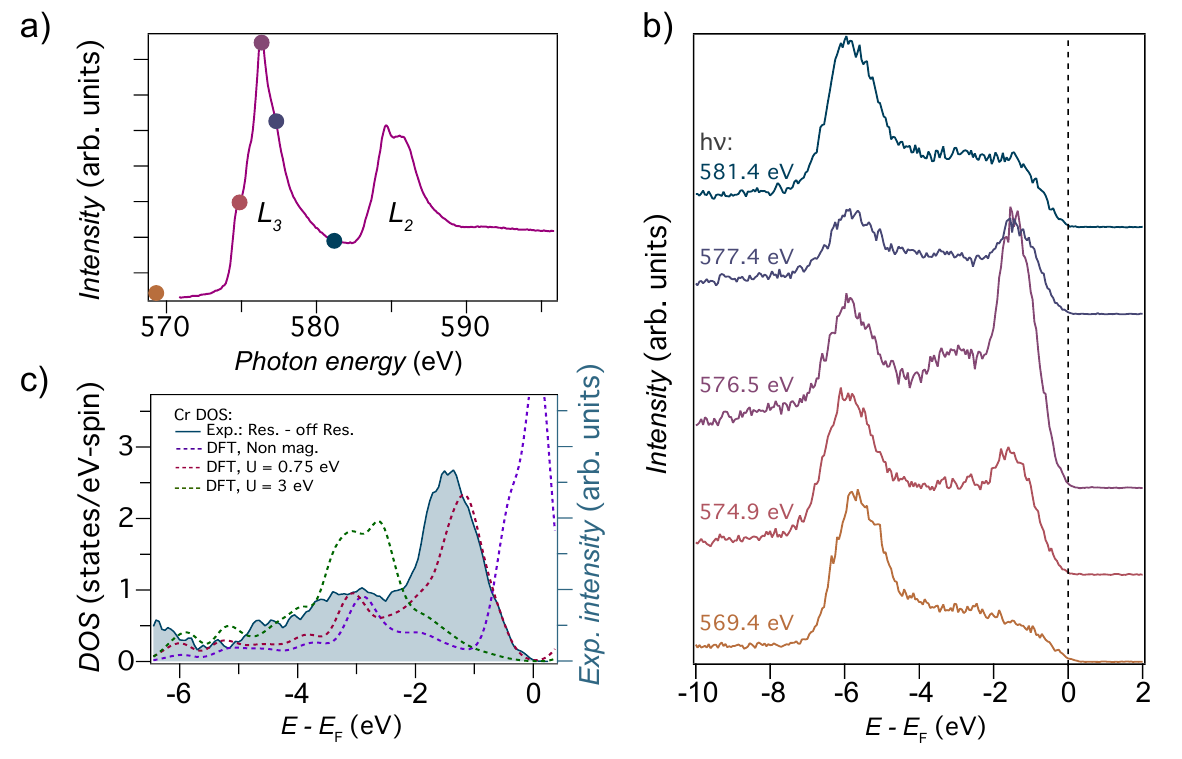}
	\caption{(a) X-ray absorption spectroscopy at the Cr $L_{2,3}$-edge, indicating a dominant Cr$^{3+}$ valence. (b) Angle-integrated resonant photoemission measurements of the valence band electronic structure, measured at the photon energies marked by the points in (a) which span across the $L_3$ edge. (c) The Cr partial density of states extracted experimentally as the difference of the on- and off-resonant PES spectra ($h\nu=576.5$~eV and $569.4$~eV, respectively (solid blue line with shading). The dashed lines show the corresponding Cr partial-DOS as determined by non-magnetic DFT, and by LDA+$U$ calculations for $U=0.75$~eV and 3~eV, respectively.}
	\label{PES_figure}
\end{figure}
\begin{figure}
	\centering
	\includegraphics[width=0.8\columnwidth]{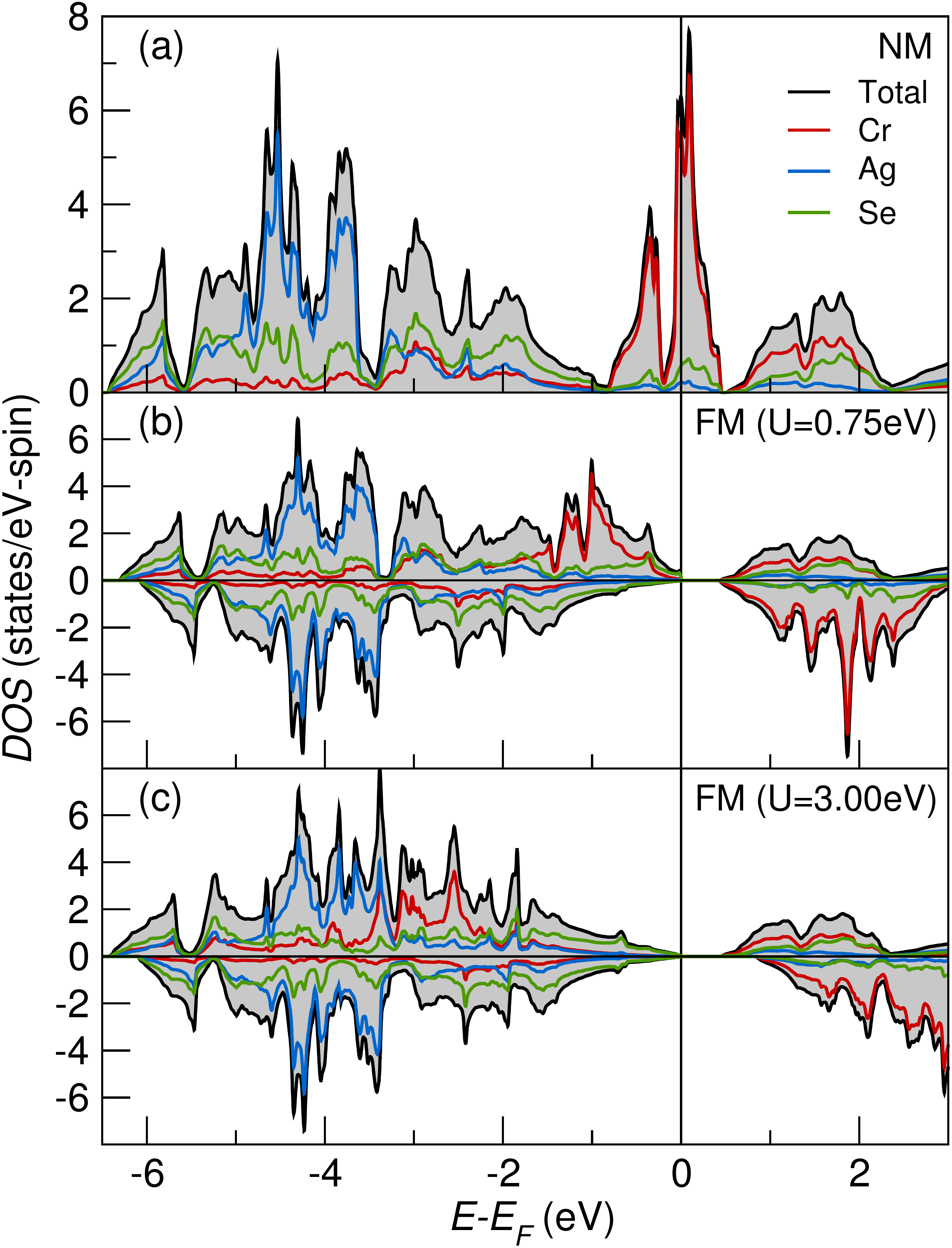}
	\caption{Total and partial electronic densities of states of AgCrSe$_2$. (a) Non-magnetic metallic solution with dominating Cr $3d$ states at $E_F$ (b) Ferromagnetic insulating state including spin orbit coupling for $U=0.75$~eV and (c) $U=3$~eV }
	\label{DFT_figure}
\end{figure}

Fig.~\ref{PES_figure}a shows XAS measurements of the Cr $L_{2,3}$-edge. Its shape is characteristic of a dominant Cr$^{3+}$ configuration \cite{meyers_zhang-rice_2013,sunko_probing_2020,noh_direct_2015}, and is similar to the XAS measured from PdCrO$_2$ \cite{sunko_probing_2020,noh_direct_2015}. A Cr $3d^3$ electronic configuration would thus be expected to first order, with a half-filled $t_{2g}$ shell. In the absence of strong electronic correlations or the formation of local magnetic moments, this would lead to a metallic ground state. Such itinerant Cr states are indeed obtained from our non-magnetic density-functional calculations within the local density approximation (Fig.~\ref{PES_figure}c and Fig.~\ref{DFT_figure}). In contrast, our measurements of the valence DOS (Fig.~\ref{PES_figure}b) indicate only a weak step at the Fermi level, with the majority of the occupied DOS located at higher binding energy.

Our photoemission measurements were performed with photon energies spanning the Cr $L_3$ X-ray absorption edge [Fig.~\ref{PES_figure}(b)]. We observe a marked increase in the spectral weight of the peak in the DOS centered at $\sim\!1.5$~eV binding energy when the photon energy is tuned into resonance with the corresponding Cr $2p$ -- $3d$ transition. This indicates that this feature has a dominantly Cr-derived character, whose spectral weight becomes enhanced `on-resonance'. To investigate this in more detail, we take the difference of the photoemission spectra measured on and just below the resonance energy, allowing us to extract an experimental measure of the Cr partial-DOS [Fig.~\ref{PES_figure}(c)]. This confirms that there is negligible Cr-DOS at the Fermi level. As our measurements are performed well above the magnetic ordering temperature of AgCrSe$_2$ (see below), this provides initial evidence for the persistence of a well-defined local moment above $T_N$.

In PdCrO$_2$, which has similar Cr-derived local moments persisting above $T_N$, the electronic structure of the CrO$_2$ subsystem is best described as Mott insulating \cite{sunko_probing_2020,lechermann_hidden_2018}. In AgCrSe$_2$, however, the larger Se orbitals may lead to increased hybridization. Indeed, a lack of definition in the $L_3$-edge XAS pre-peak structure as compared to PdCrO$_2$ is consistent with an increased hybridization with the ligand; similar to that observed in other Cr-based chalcogenide and halide magnets such as Cr$_2$Ge$_2$Te$_6$ \cite{watson_direct_2020} and CrI$_3$ \cite{frisk_magnetic_2018}. This, in turn, would weaken the electronic correlations. To quantify this, we compare our experimental Cr partial-DOS with that calculated from DFT (Fig.~\ref{PES_figure}c and Fig.~\ref{DFT_figure}). A clear disagreement is found with the non-magnetic DFT calculations, where a large Cr-DOS is obtained at the Fermi level, unlike that obtained experimentally. We thus compare with magnetic LDA+$U$ calculations, where the strength of the Hubbard-$U$ parameter is varied. Taking a value of $U=3$~eV, typical for Cr-based oxides \cite{janson_2013_Cr,janson_2014_Cr}, we find that the Cr-DOS is peaked at too high a binding energy as compared to the experiment. Moreover, rather than having a single dominant peak as for the $U=3$~eV calculations, the experimental partial DOS has a double-peak structure with spectral weight distributed throughout the valence band. Indeed, much better agreement with both the locations and overall shape of the experimental DOS is obtained for calculations using a low $U$ value of 0.75~eV. This points to a much more weakly correlated state as compared to e.g.\, PdCrO$_2$, where the Cr-derived states in AgCrSe$_2$ become more strongly hybridised with the ligand states. This can be expected to have an important influence on the magnetic ordering in AgCrSe$_2$, as discussed in detail below.

Carrying out DFT calculations for different settings with respect to magnetism or electron correlation can provide a deeper understanding for different features in the underlying electronic structure, in particular the influence of the different constituents. Figure \ref{DFT_figure} shows the total and partial DOS's resulting from such calculations. Panel \ref{DFT_figure}(a) shows a nonmagnetic calculation, resulting in a metallic solution with dominating, half filled $t_{2g}$ Cr $3d$ states (in terms of the local CrSe$_6$ octahedron) at the Fermi level $E_F$. With respect to the Cr states, this solution is in clear contrast to the Cr partial DOS determined from photoemission experiments discussed above, and also experiments which find the compound to be a doped semiconductor with a small charge carrier density and magnetic moments on the Cr site. However, the position of the Se and Ag states in the valence band meets the expectations from related compounds: a rather narrow Ag dominated band complex around $-4$~eV and a broad Se contribution spread throughout the valence band due to hybridization with both Cr and Ag.

To reflect more of the experimental properties, including the local Cr moment should provide considerable improvement, leading to a split of the Cr $t_{2g}$ states with full spin polarization and the opening of a band gap. To approximate the experimentally observed complex noncollinear in-plane magnetic structure with a long propagation vector (see below), we use ferromagnetic (FM) order. As expected from a local moment picture, we find the system to be insulating with a gap of about 12~meV (not shown). The experimentally observed metallic conductivity with a small charge carrier concentration is related to a inherent small Ag off-stoichiometry of our crystals and the related self doping.

However, compared to the PES data, the FM calculation yield still a too high Cr contribution near $E_F$. A likely explanation for this discrepancy are Coulomb correlations in the Cr $3d$ shell. Since such correlation effects are insufficiently described by standard DFT,  we apply FM LDA+$U$ calculations to take them into account in a mean field way. As discussed above, a rather moderate value of $U=0.75$~eV  fits the PES data best [see Figs.~\ref{DFT_figure}(b) and \ref{PES_figure}(c)]. Besides an increase of the band gap, the respective small down shift of the Cr $3d$ states leads to a mixture of Cr $3d$ and Se $4p$ states near $E_F$ with a larger weight for the Se states. A larger $U$ value of $3$~eV, typical for Cr-based oxides \cite{janson_2013_Cr,janson_2014_Cr}, moves the Cr $3d$ states too far down in energy (see Figs.~\ref{DFT_figure}(c) and \ref{PES_figure}(c)). In such a scenario, the states near $E_F$ are heavily dominated by Se $4p$ contributions [see Fig.~\ref{DFT_figure}(c)], resulting in sizeably increased dispersion of the respective bands \cite{Kim}.

Having adjusted $U$ to fit the PES data, we further analyse the resulting band structure. In a tight binding model for the $t_{2g}$ states, we find a dominant in-plane dispersion with about equal contribution from first- and third-nearest neighbors. The inclusion of spin orbit coupling (SOC) leads to a significant additional split of the order of 300 meV (on the $\Gamma$-Z line) for the top of the valence band. The details will be published elsewhere \cite{Kim}.

\subsection{Magnetic susceptibility and magnetization} \label{SEC:chi}

\begin{figure}[bt!]
\includegraphics[clip,width=\columnwidth]{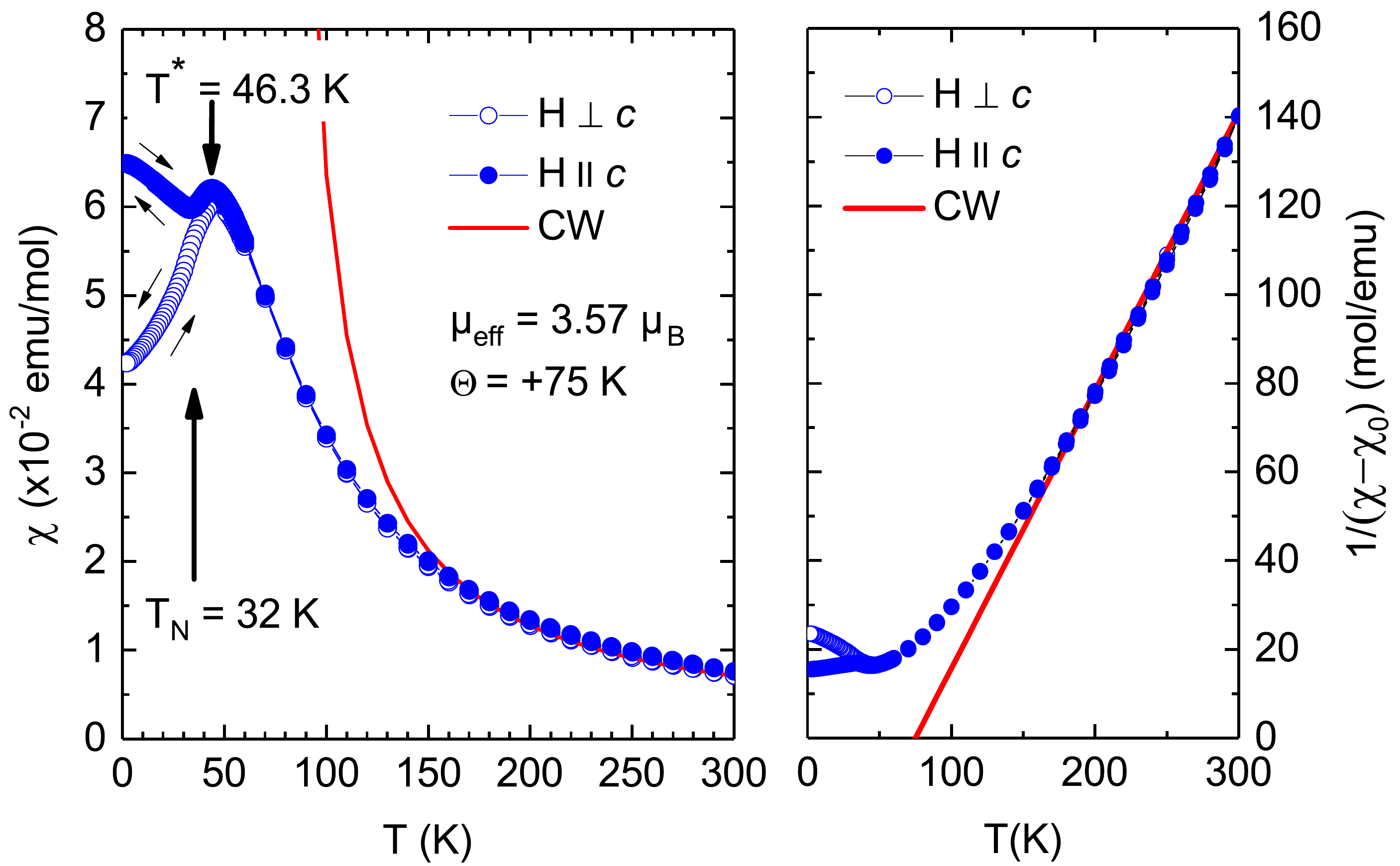}
\caption{Temperature dependence of the magnetic susceptibility ($\chi_\parallel$ and $\chi_\perp$) of AgCrSe$_{2}$ (MS1212) measured at $\mu_{0}H=1$~T (Left).  The solid line corresponds to a Curie-Weiss fit. Each measurement was first performed in zero-field-cooled (zfc) and then in field-cooled (fc) mode.  Inverse of the susceptibility as a function of temperature (Right. Note that a paramagnetic contribution for $H \parallel c$ of $\chi_{0, \parallel} = 0.0005$~emu/mol was required for a proper Curie-Weiss fit (CW) at high temperatures. Most likely this stems from a van Vleck contribution ($\chi_{VV}$) (estimated from high field measurements (Fig.~\ref{MparaxH})) and a small and negative diamagnetic contribution from the capton sample holder $\chi_{0,\parallel} = \chi_{VV,\parallel}+\chi_{dia}$.}
\label{ChixT}
\end{figure}

\begin{figure}[bt!]
\includegraphics[clip,width=0.9\columnwidth]{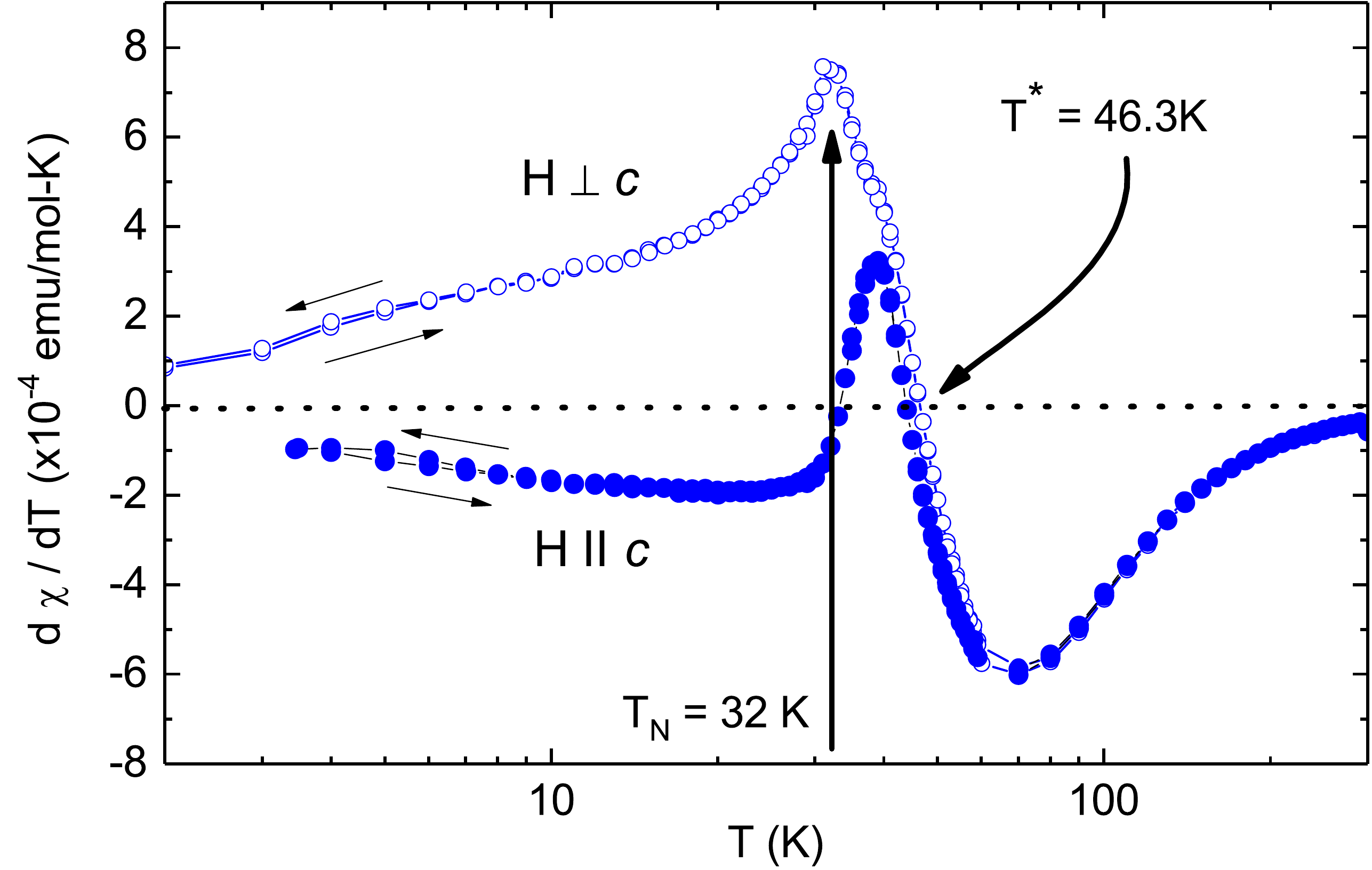}
\caption{First temperature derivative of the magnetic susceptibility ($\chi_\parallel$ and $\chi_\perp$) of AgCrSe$_{2}$ (MS1212) as a function of temperature. The maximum in susceptibility at $T^* = 46.3$~K (Fig.~\ref{ChixT}) is given by the high temperature zero crossing of the derivative. The ordering temperature of $T_N = 32$~K is indicated by a maximum in the derivative for $H \perp c$ and the second zero crossing in the derivative for $H \parallel c$.}
\label{DChixT}
\end{figure}

Figure~\ref{ChixT} shows the magnetic susceptibility of AgCrSe$_{2}$ as a function of temperature in a magnetic field of 1~T applied in the ($a$,$b$) - plane $H  \perp c$ and in the $c$ direction $H \parallel c$. Below the ordering temperature of $T_{N}\approx32$~K a strong magnetic anisotropy is observed. Above $T_{N}$, the magnetic anisotropy disappears and a pronounced maximum at $T^{*}= 46.3$~K is found in $\chi_{\perp}(T)$ and $\chi_{\parallel}(T)$.  Above 150~K, $\chi_{\perp}(T)$ and $\chi_{\parallel}(T)$ can be well fitted by a Curie-Weiss (CW) law, yielding a positive Weiss temperature of $\theta = +75$~K and an effective moment of $\mu_{\rm eff} = 3.57$~$\mu_{\rm B}/{\rm Cr}$. This value of $\mu_{\rm eff}$ is slightly smaller than the theoretical prediction for the spin-only value of trivalent Cr (3$d^{3}$) with $S=3/2$ and $\mu^{\rm theo}_{\rm eff}=3.87$~$\mu_{\rm B}$. A maximum at about 50~K, a Weiss temperature around $70$~K and an effective moment of about $\mu_{\rm eff} = 3.52$~$\mu_{\rm B}/{\rm Cr}$ have been also reported in earlier studies on polycrystalline powdered samples of AgCrSe$_{2}$ \cite{Bongers,Gautam}. However, upon lowering the temperature the susceptibility measurements on the powder show a structure-less drop in susceptibility after passing through the maximum at about 50~K. To our knowledge, so far neither directional measurements on single crystals nor detailed studies on the field dependence of the magnetization were reported for the Cr-sulfide or the Cr-selenide delafossite systems in the literature. Thorough early neutron diffraction experiments on polycrystalline AgCrSe$_{2}$ show a complex planar cycloidal spin arrangement in the $(1,1,0)$ direction with a large periodicity of about 164~\AA\ in zero field \cite{Engelsman}. In the $c$ direction the FM planes are antiferromagnetically coupled which leads to a global AFM order. This is a distinct difference to the $R\bar3m$ Cr-oxygen delafossites, such as AgCrO$_{2}$, CuCrO$_{2}$, or PdCrO$_{2}$, where a planar $120^{\circ}$ order has been established \cite{Vandenberg,Seki,Billington}. In contrast to the Cr-sulfide or Cr-selenide systems, the Cr oxygen-based delafossites always possess large negative Weiss temperatures  ($\theta = -160$~K for AgCrO$_{2}$, $\theta =-211$~K for CuCrO$_{2}$, and $\theta =-500$~K for PdCrO$_{2}$  \cite{Oohara,Kimura,Maeno}), indicating a predominant AFM exchange. For non-oxygen Cr delafossites the exchange becomes quite complex and studies on polycrystalline samples evidence positive Weiss temperatures for AgCrSe$_{2}$, NaCrSe$_{2}$, CuCrSe$_{2}$ NaCrS$_{2}$ and KCrS$_{2}$, as well as negative Weiss temperatures for AgCrS$_{2}$, CuCrS$_{2}$ and LiCrS$_{2}$ \cite{Bongers,Engelsman,Vandenberg,Kobayashi}.

\begin{figure}[bt!]
\includegraphics[clip,width=\columnwidth]{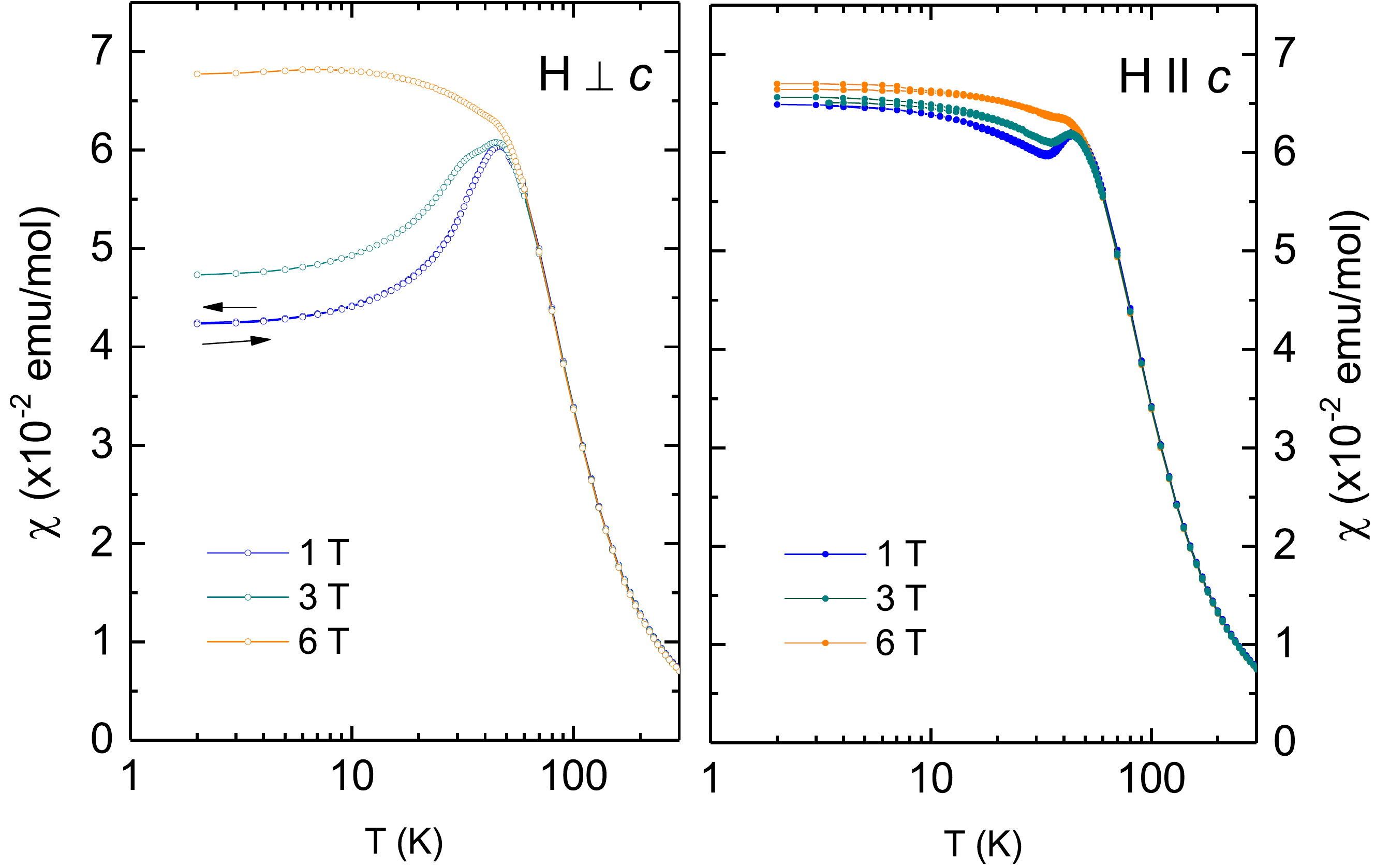}
\caption{Temperature dependence of the magnetic susceptibility of AgCrSe$_{2}$ (MS1212) measured at various magnetic fields for $H  \perp c$ (left) and  $H \parallel c$ (right). All measurements are performed in zfc/fc mode.}
\label{ChixT_sevH}
\end{figure}

\begin{figure}[bt!]
\includegraphics[clip,width=0.9\columnwidth]{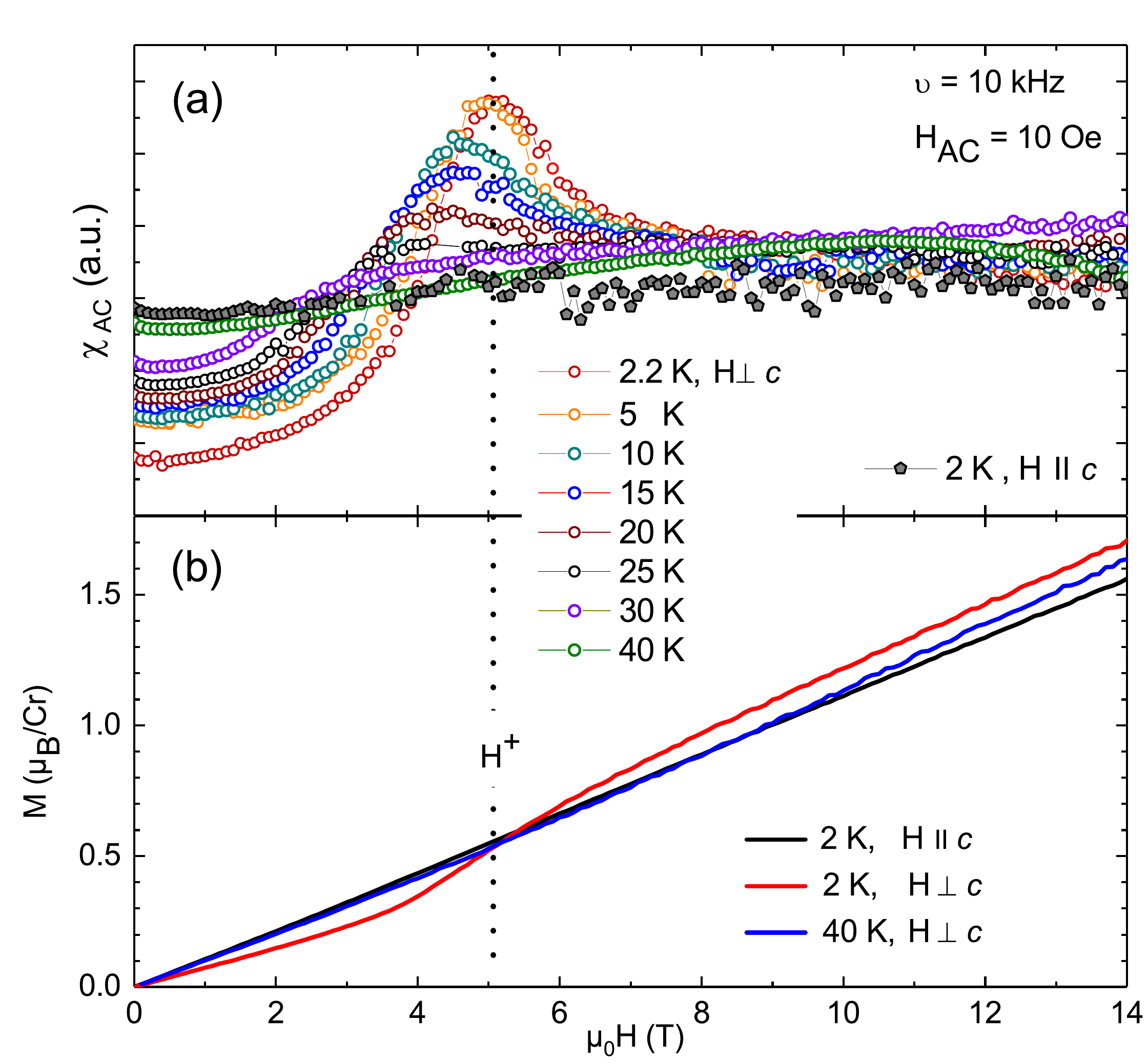}
\caption{AC susceptibility (a) and DC magnetization (b) versus field for $H \parallel  c$ and $H \perp c$ at various temperatures as indicated. The DC and AC magnetization were measured following a zfc protocol and by increasing the field from zero to 14~T followed by warming up to 200~K between the individual cycles. }
\label{AC_Chi}
\end{figure}

We use the first temperature derivative of the magnetic susceptibility $d\chi/dT$ to determine the magnetic ordering temperature (see Fig.~\ref{DChixT}). For fields in the ($a$,$b$) - plane we find a clear change in the slope of the $\chi_{\perp}(T)$ curve which gives a pronounced peak at $T_{N}=32$~K, whereas in $c$ direction a zero crossing of the derivative of $\chi_{\parallel}(T)$ indicates the magnetic order at the same temperature within the measurement uncertainties. The position of the isotropic maximum at 46.3~K in $\chi_{\perp}(T)$ and $\chi_{\parallel}(T)$ is defined by the second zero crossing in the first derivative. We note that the measurements were performed in zero-field-cooled (zfc) and field-cooled (fc) mode and that no branching could be observed. This evidences that there is no hysteretic behavior due to short-range order, domains, or other glassy-like effects. Furthermore, this behavior is reproducible across many growth series of  AgCrSe$_{2}$ crystals (see equivalent measurements on MS1257 crystal in the Supplemental Material \cite{SM}). The Weiss temperature of $\theta = +75$~K and the ordering temperature of $T_{N}=32$~K give a frustration parameter of $f = \theta/T_{N}=2.3$. At first sight and compared to the oxy-chromium delafossites with large negative Weiss temperatures and predominant in-plane AFM interaction this seems to contradict a strong frustration. However, we point out that in the AgCrSe$_{2}$ we have AFM and FM interactions in the plane (see Sec.\ \ref{SEC:theory}).

\begin{figure}[bt!]
\includegraphics[clip,width=0.8\columnwidth]{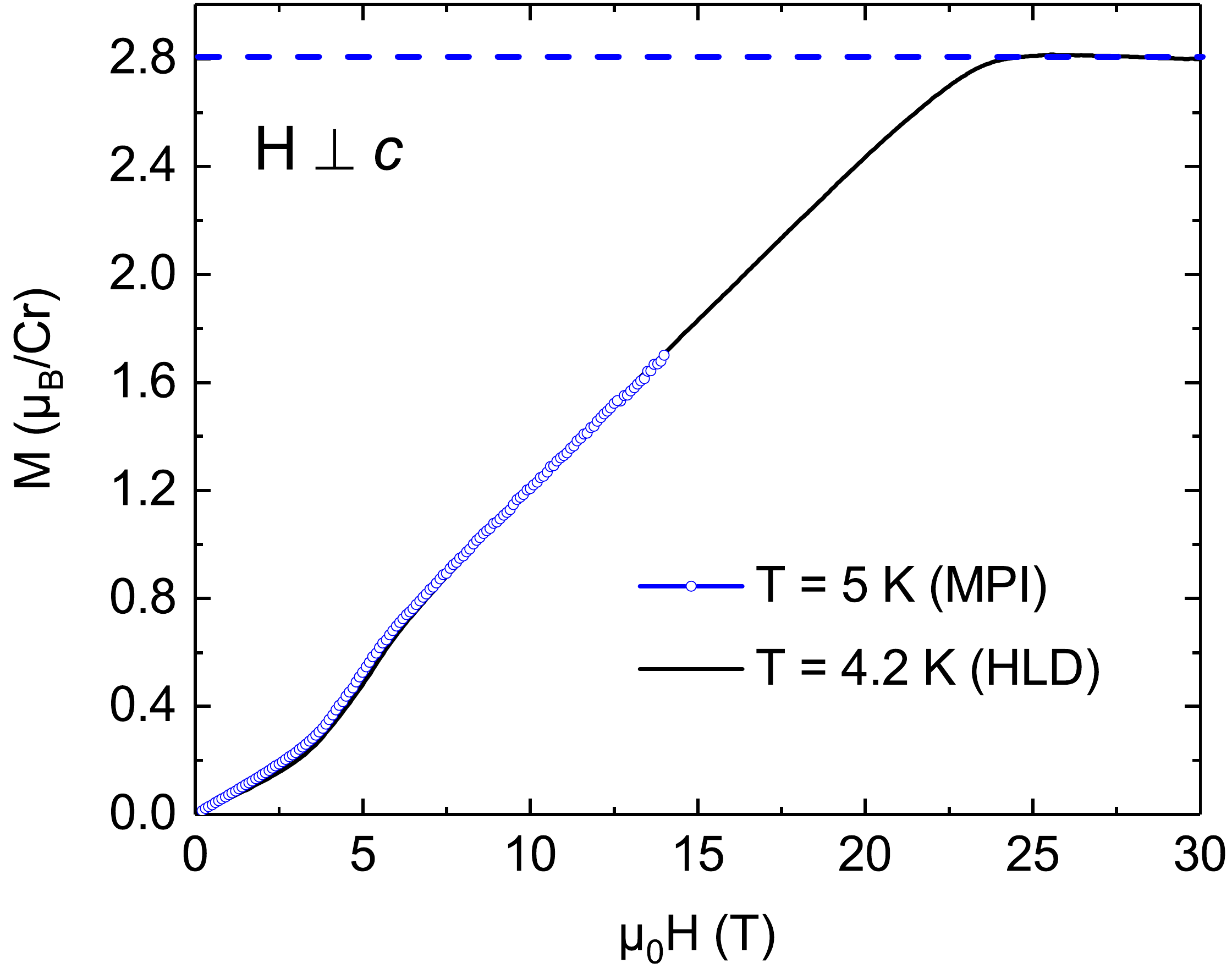}
\caption{Magnetization versus field for $H \perp c$. Open symbols represent data from measurements in static fields in a PPMS (at the MPI) and solid lines represent high field magnetization data recorded in pulsed fields at the HLD.}
\label{MperpxH}

\end{figure}

The effect of an increase of the magnetic field on $\chi_{\perp}(T)$ and $\chi_{\parallel}(T)$ is demonstrated in Fig.~\ref{ChixT_sevH}. The maximum at about $T^{*}= 46$~K remains rather robust against increasing fields for both directions. There is almost no field effect on $\chi_\perp(T)$ and $\chi_{\parallel}(T)$ toward higher temperatures. In contrast, there are dissimilar field dependencies below $T^{*}$: for $H \perp c$ the susceptibility $\chi_\perp(T)$ increases strongly with field, whereas for $H \parallel c$ there is only a rather small increase detectable. This is a first evidence for a field-induced metamagnetic transition in the $H\perp c$ direction. We further investigated this effect by AC susceptibility and magnetization measurements up to 14~T. Here the field-induced transition at $H^{+}(T)$ becomes evident by the evolution of a pronounced peak in the AC susceptibility [Fig.~\ref{AC_Chi}(a)] and
a slope change of the magnetization $M(H)$ [Fig.~\ref{AC_Chi}(b)]. For $H \parallel c$ these features are absent.

\begin{figure}[bt!]
\includegraphics[clip,width=0.7\columnwidth]{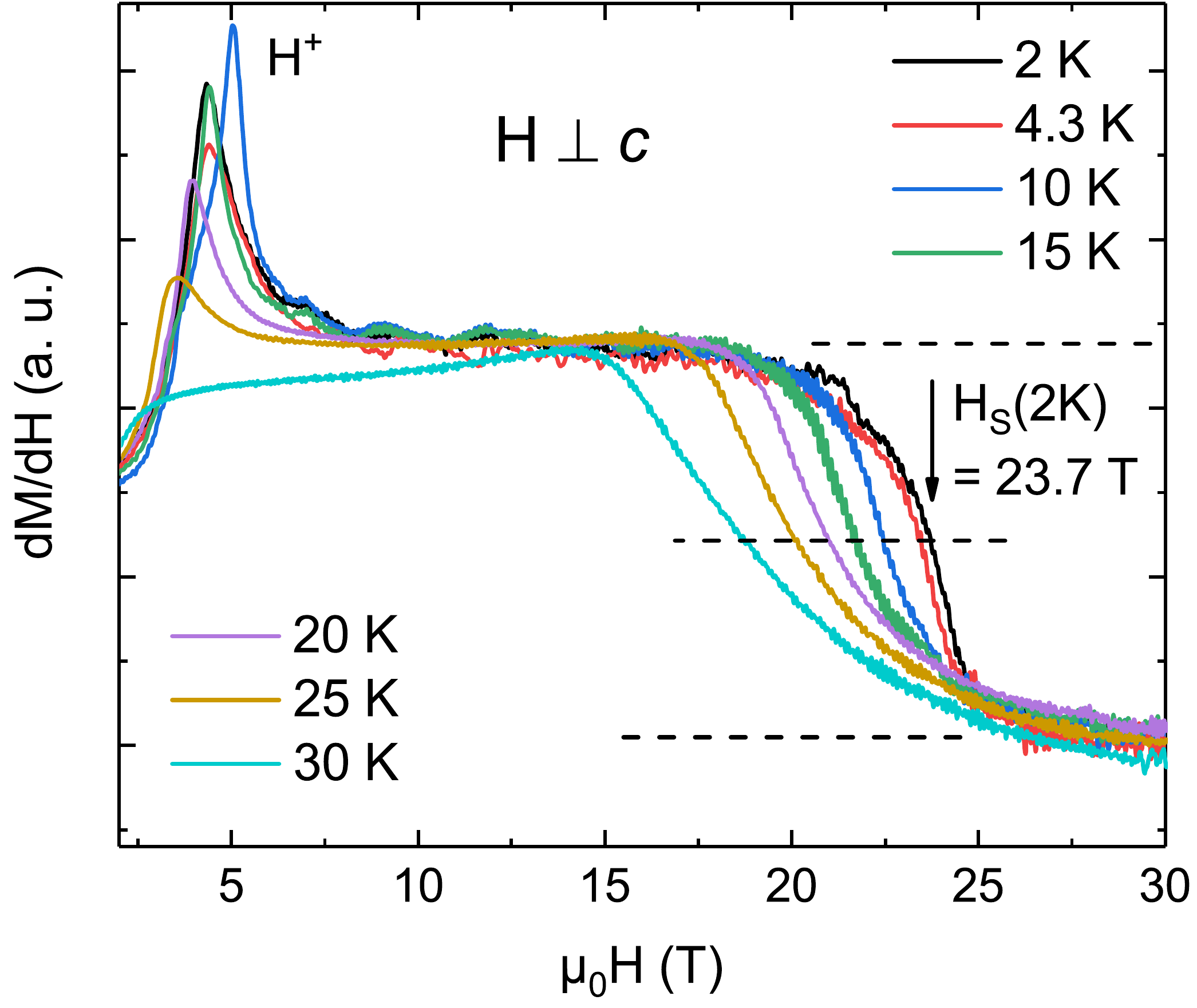}
\caption{First field-derivatives of the magnetization versus field for $H \perp c$ at various temperatures. The metamagnetic transition field $H^{+}$ is defined by a pronounced peak, whereas the onset of saturation at $H_S$ is given by a step in the derivatives.}
\label{DMperpxH}
\end{figure}

\begin{figure}[bt!]
\includegraphics[clip,width=1\columnwidth]{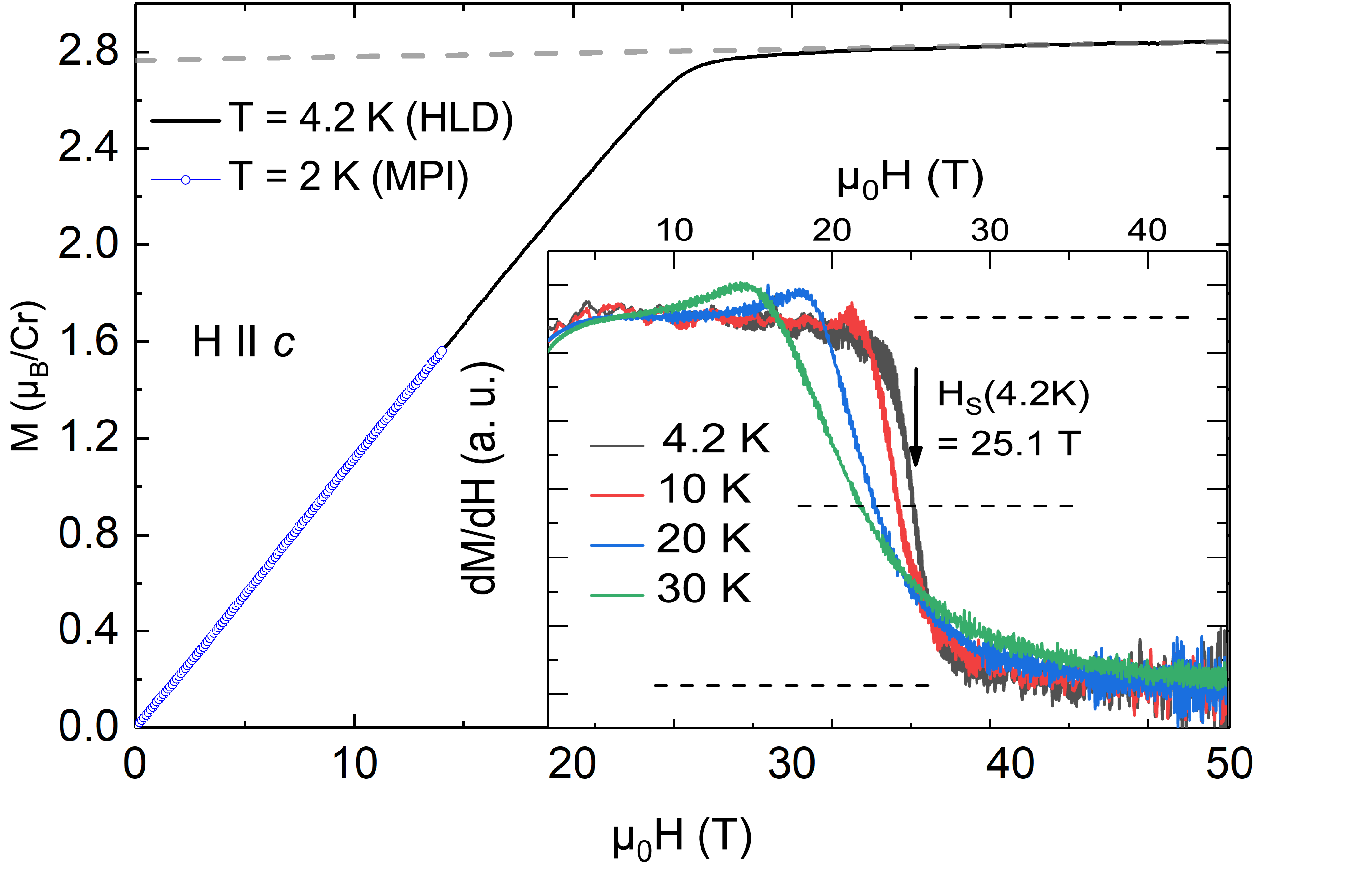}
\caption{Magnetization versus field for  $H \parallel c$. Open symbols represent data from measurements in static fields in a PPMS and solid lines represent high field magnetization data obtained in pulsed fields at the HLD. The dashed line corresponds to the linear in field van Vleck contribution ($M_{VV}= H \chi_{VV\parallel}$) above 40~T with  $\chi_{VV\parallel} = 0.0008$~emu/mol. The inset shows the first field derivative of the magnetization versus field for $H \parallel c$ at various temperatures. The saturation field $H_S$ is given by a step in the derivatives. }
\label{MparaxH}
\end{figure}

The magnetization $M(H)$ was measured for $H\perp c$ and $H\parallel c$, shown in Figs.~\ref{MperpxH} and \ref{MparaxH}, respectively, in fields up to 14~T, using  a PPMS and in pulsed magnetic fields up to 50~T at the HLD. Data on a polycrystalline sample can be found in the Supplemental Material \cite{SM}. For $H \perp c$ a field-induced metamagnetic transition at $H^{+}(\rm 5~K) = 5$~T in $M(H)$ is followed by a saturation of the magnetization at $M_s = 2.8$~$\mu_{\rm B}/{\rm Cr}$ above $H_s = 23.7$~T (Figs.~\ref{MperpxH} and \ref{DMperpxH}). Figure~\ref{DMperpxH} shows the first derivative of the magnetization $dM/dH$ at various temperatures.  $H^{+}(T)$ is defined by the peak in the derivative, in good agreement with the AC susceptibility results. The crossover to saturation at high fields is broadened toward higher temperatures. We use the half height of the high-field step in the first derivative as a measure for the transition to saturation at $H_s$. From the ESR an isotropic saturation magnetization of about $M_{S}={g}J=3$~$\mu_{\rm B}/{\rm Cr}$ is expected for $J=3/2$ and $g=2$ (see Sec.\ \ref{SEC:ESR} on the ESR results). The experimentally determined saturation magnetization of $2.8~\mu_{\rm B}/{\rm Cr}$ does not fully reach the value predicted by the ESR. Possible reasons for that might be an orbital moment contribution, in addition to the spin-only moment, or shielding effects from itinerant conduction electrons with admixed Cr-$3d^3$ character.

\subsection{Specific heat and thermal expansion}\label{SEC:heat}

\begin{figure}[bt!]
\includegraphics[clip,width=0.8\columnwidth]{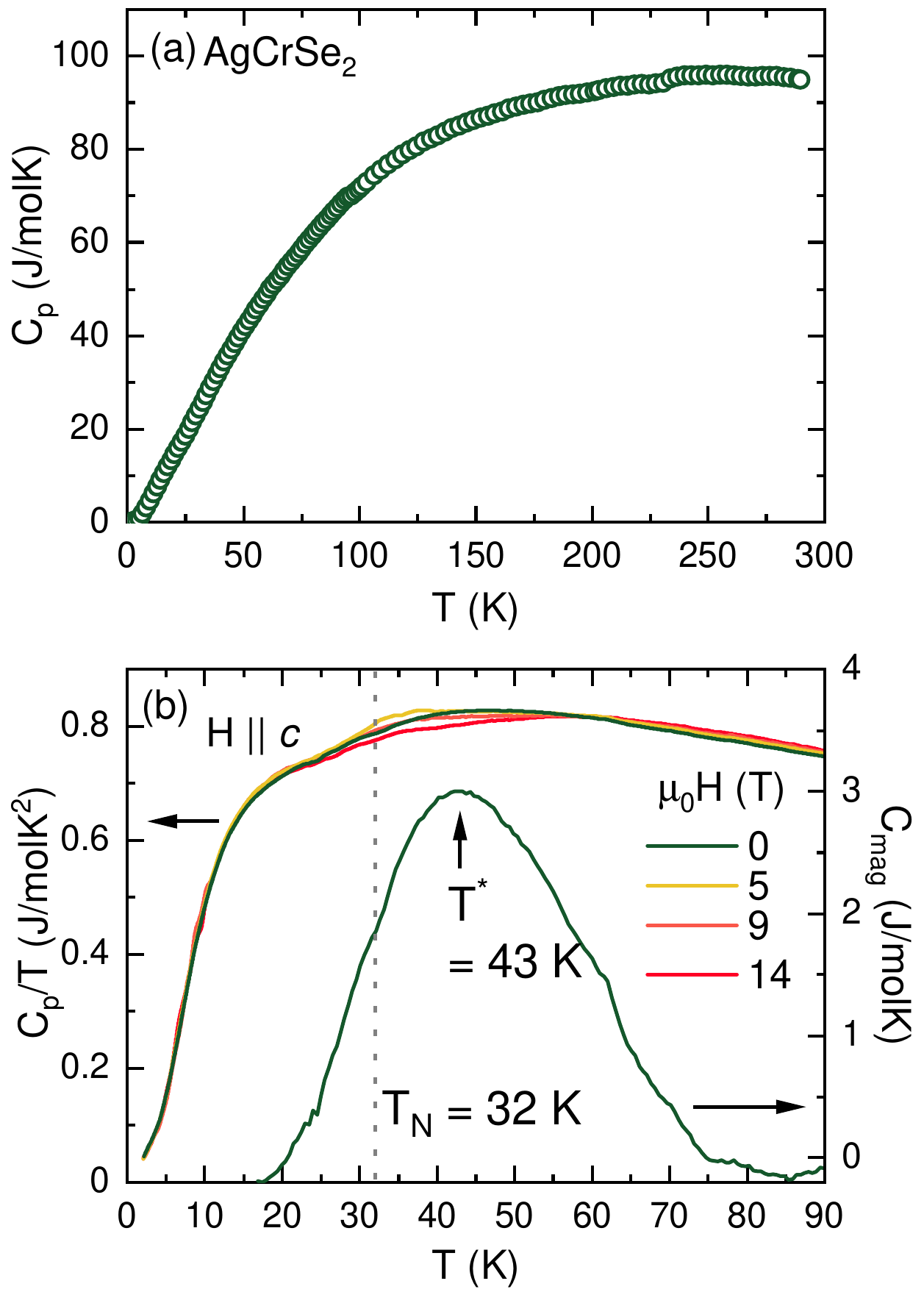}
\caption{(a) Temperature dependence of the zero-field specific heat $C_{\rm p}$ of an AgCrSe$_{2}$ single crystal. (b) $C_{\rm p}/T$ vs $T$ for various fields for $H  \parallel c$ (left axis) and magnetic specific heat $C_{\rm mag}$ at zero field as function of temperature (right axis).  The maximum in $C_{\rm mag}$ at $T^* = 43$~K is clearly visible and the maximum value $C_{\rm mag}(\rm 43~K)$ corresponds to 0.36~R.}
	\label{Cp}
\end{figure}

\begin{figure}[bt!]
\includegraphics[clip,width=0.8\columnwidth]{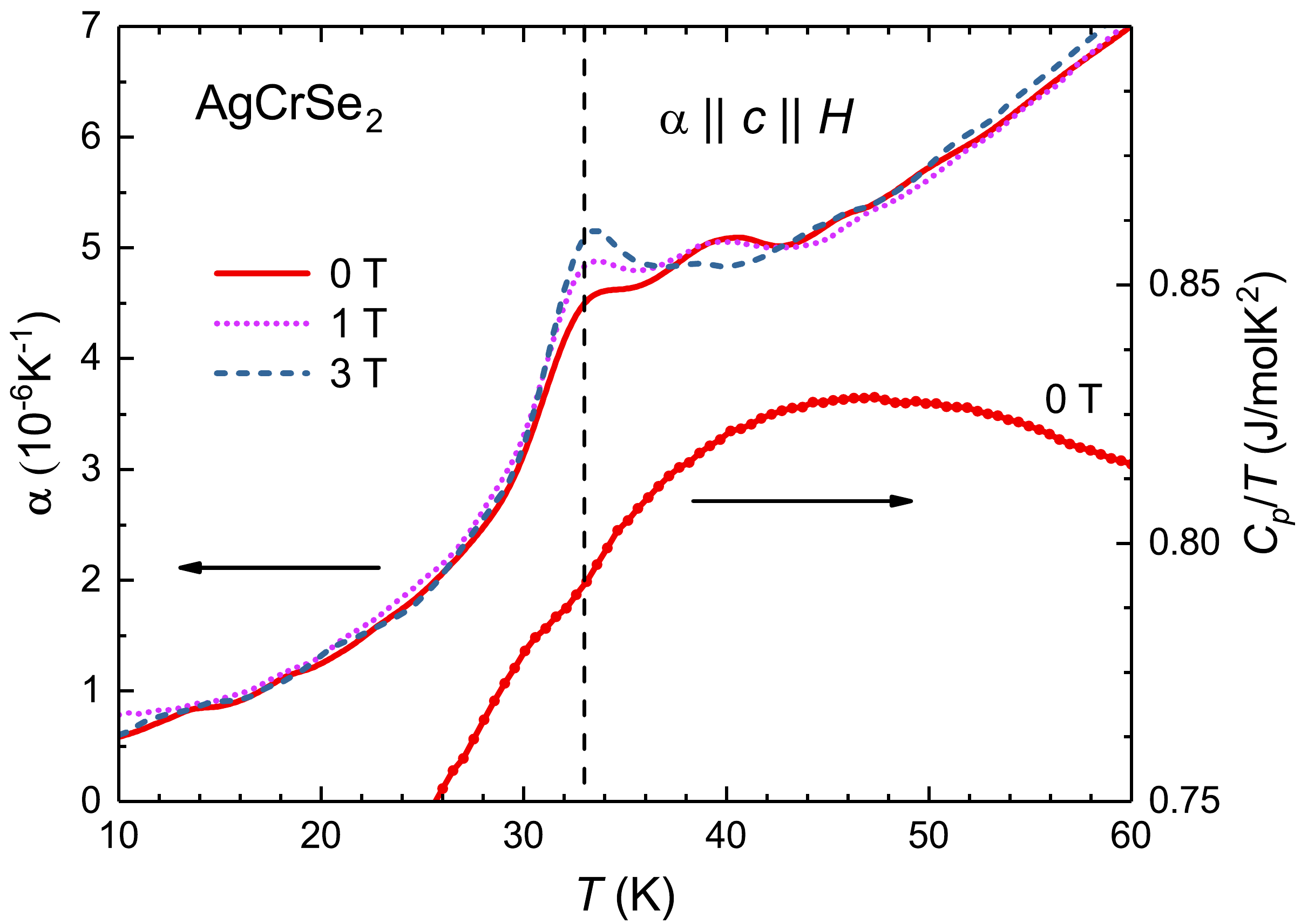}
\caption{Thermal expansion coefficient $\alpha$ measured in $c$ direction as function of temperature (left scale) together with $C_{\rm p}/T$ (expanded view) (right scale). The dashed vertical line marks the largest anomaly in the expansion at T$_{N} \approx $ 33~K.}
\label{Thermal+Cp}
\end{figure}

Figure~\ref{Cp}(a) presents the specific heat ($C_{p}$) of AgCrSe$_{2}$ as a function of temperature in the temperature range from 2 to 280~K. In strong contrast to oxygen-based Cr-delafossites, such as  CuCrO$_{2}$ \cite{CuCrO2} and PdCrO$_{2}$ \cite{Maeno}, no clear $\lambda$-type anomaly indicating a magnetic phase transition can be observed. The absence of a strong signature might be due to the complex cycloidal magnetic order triggered by the non-centrosymmetric structure which promotes an extended spin entanglement (about 164~\AA~ in (1,1,0) direction \cite{Engelsman}), the frustration due to competing AFM and FM exchange, short range correlations, and emergent fluctuations. Only a tiny anomaly at the phase transition at about 33~K can be resolved in the specific heat data (see Fig.~\ref{Thermal+Cp}). A broad maximum can be found in the zero-field magnetic specific heat ($C_{mag}$) as a function of temperature [see Fig.~\ref{Cp}(b)]. The presence of this broad peak centered around $T^* = 43$~K is in agreement with the maximum observed in the magnetic susceptibility (see Fig.~\ref{ChixT}) and once again points at the presence of strong frustration in AgCrSe$_{2}$. For planar isotropic triangular lattices the theory predicts this maximum with an absolute value of $C_{mag}(T^*) = 0.4 {\rm R}$ \cite{schmidt2,Bernu}. Even so most of the model calculations are performed for $S=1/2$ ions our experimental value of $C_{mag}(T^*) \approx 0.36 {\rm R}$ for a $S=3/2$ Cr lattice is close to this prediction.  The application of magnetic fields leads to a broadening of this peak as can be seen in Fig.~\ref{Cp}(b). To obtain the magnetic part of the specific heat $C_{mag}$, we subtract the phonon and electronic contributions determined by fitting a combination of Einstein, Debye, and linear terms to the data. The details are described in the Supplemental Material \cite{SM}. In zero field the integration of $C_{mag}/T$ from 2 to 70~K yields a reduced magnetic entropy of about 18\% of ${\rm R}\ln4$. This reduced entropy might be the fingerprint of the complex cycloidal order and evolving fluctuations toward long-range order as evidenced by  neutron diffraction (see Sec.\ \ref{SEC:Neutron}). In low dimensional strongly frustrated quantum magnets it is frequently found that the entropy is reduced due to strong frustration and associated fluctuations of the moments. Usually the entropy recovers and the signature of the order sharpens upon the application of magnetic field due to quenching AFM fluctuations and the polarization of the moments \cite{Ramirez}. Here we observe a broadening and a shift of the magnetic contribution to higher temperatures upon increasing field, which points at the importance of the FM exchange component [see Figure~\ref{Cp}(b)]. Due to the broad feature and the large nonmagnetic phonon contribution we cannot reliably determine  $C_{mag}(T,H)$ from the experimental data set in magnetic fields.

In contrast to AgCrSe$_{2}$, other non oxygen-based Cr triangular lattice systems within the $R3m$ symmetry group (AgCrS$_{2}$, CuCrS$_{2}$) undergo an additional structural phase transition to a lower symmetry group at low temperatures, and therefore show a pronounced anomaly associated to the simultaneous structural and magnetic order in the low temperature heat capacity \cite{Vandenberg,ushakov,AgCrS2}. Furthermore, they have negative Weiss temperatures, i.e.\ they have predominant AFM exchange in the plane. The specific heat of CuCrSe$_{2}$ is more similar to our results \cite{Tewari} and the system shows no symmetry reduction towards low temperatures. Here one also finds a rather broad peak in the magnetic specific heat and only a tiny anomaly at the magnetic ordering temperature of about 50~K and a reduced magnetic entropy of about 75\% of ${\rm R}\ln4$. Furthermore, CuCrSe$_{2}$ is metallic, undergoes a transition from $R\bar3m$ to $R3m$ at 365 K and has a positive Weiss temperature ($\theta = +5$~K) indicating some in-plane FM exchange making the system a close relative of AgCrSe$_{2}$\cite{Gagor}.

To gain further insight into the nature of the phase transition, thermal expansion was measured in the temperature range from 80~K down to about 6~K.  Here, the linear thermal expansion coefficient $\alpha$ is defined as
\begin{equation}
	\alpha=\frac{1}{L}\frac{dL}{dT},
	\label{alpha_cp}
\end{equation}
where $L$ is the length of the sample (in this case along the crystallographic $c$ direction).  The values of $\alpha$ obtained in fields $\mu_0H = 0$, 1, and 3~T are presented in Fig.~\ref{Thermal+Cp}.  The data at zero field clearly confirm a transition at $T\approx 33$~K, which is not of first order.  The latter may explain the only small hump seen in the $C_p/T$ data at this temperature, plotted for comparison in Fig.~\ref{Thermal+Cp}  Great care was taken to accurately calculate $\alpha$ from the measured sample length; in fact, a second sample was measured to confirm the broad hump feature right above the transition temperature.  Moreover, the thermal expansion measured in small magnetic fields applied parallel to crystallographic $c$ direction corroborates the thermodynamic nature of the transition.  As indicated by the dashed line in Fig.~\ref{Thermal+Cp}, the transition is observed at nearly constant temperature for $\alpha \parallel c$ and within fields up to 3~T.  We note that, as a result of the platelet-like shape of the samples (see Fig.~\ref{Crystal}), thermal expansion could so far only be measured along the crystallographic $c$ direction, even though a comparison to the neutron data for other directions would be highly desirable. A comparison with the thermal expansion of Cr-sulphur systems ($A$CrS$_{2}$, $A=$ Cu, Ag, Au) clearly shows, as already mentioned in the discussion of the specific heat, a different behavior since in these systems the magnetic phase transition is accompanied by a structural phase transition \cite{Tomoya}. In contrast, our results on the thermal expansion of AgCrSe$_{2}$ clearly evidence the absence of any structural phase transition across the magnetic transition.

\subsection{Electron-spin resonance}\label{SEC:ESR}

\begin{figure}[bt!]
\includegraphics[clip,width=0.8\columnwidth]{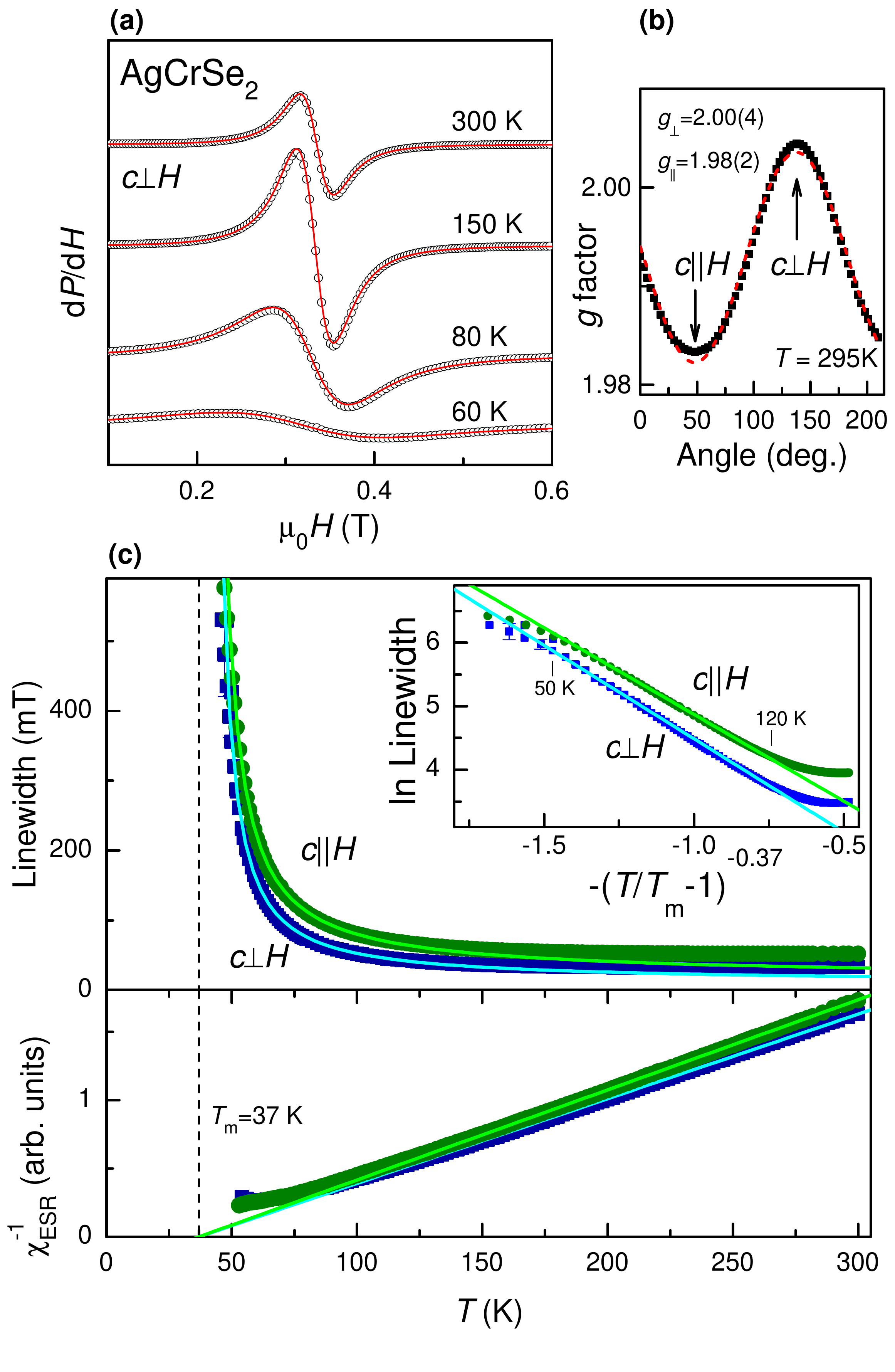}
\caption{(a) Evolution of the Cr$^{3+}$ ESR spectra with temperature, (b) $g$ factor anisotropy at 295 K, and temperature dependence of ESR linewidth and (c) inverse ESR intensity for magnetic fields $H \perp c$ and $H  \parallel  c$. Solid lines are fits to the linewidth data with a model assuming a spin relaxation via $Z_{2}$-vortices (see Refs.\ \cite{hemmida11a} and \cite{Hemmida} and main text) and a Curie-Weiss law, respectively. }
\label{ESR}
\end{figure}

Typical ESR signals at four temperatures are shown in Fig.~\ref{ESR}(a). The spectral shapes are well-defined as indicated by the Lorentzian line shape (red solid line in Fig.~\ref{ESR}(a)). The ESR $g$ factor, as determined from the resonance field at $T=295$~K, shows an uniaxial anisotropy as indicated by the red solid line in Fig.~\ref{ESR}(b) with $g_{\parallel}= 2.00(4)$ and $g_{\perp}=1.98(2)$ for the external field along the $c$ axis and in the basal plane, respectively.  The temperature dependence of the linewidth is displayed in the upper frame of Fig.~\ref{ESR}(c). Toward the magnetic ordering temperature a divergent broadening reflects the growing importance of Cr$^{3+}$ spin-correlations.
The solid lines in  Fig.~\ref{ESR}(c) depict a relaxation model assuming short range spin correlations with chiral $Z_{2}$ vortex fluctuations.
This model was previously applied to AgCrO$_{2}$ \cite{hemmida11a} and it leads to an exponential linewidth dependence on a reduced temperature scale $T/T_{\rm m}-1$ as highlighted in the inset with $T_{\rm m}$ denoting the melting temperature of the vortex lattice. Noteworthy, the same critical exponent of 0.37 as for AgCrO$_{2}$ can describe the data reasonably well suggesting a similar character of spin fluctuations in both systems. For the related system AgCrS$_{2}$ which, as already mentioned, has a structural phase transition across the magnetic phase transition, a slightly increased critical exponent of 0.42 is found \cite{Gao}. Toward high temperatures the linewidth shows a weak increase which may indicate a broadening becoming dominated by a spin relaxation toward conduction electrons. The local Cr$^{3+}$ susceptibility $\chi_{ESR}$, as determined from the ESR intensity, nicely follows a Curie-Weiss law for temperatures above $\approx 150$~K, see lower panel of Fig.~\ref{ESR}(c). The Weiss temperature agrees with $T_{\rm m} =37$~K which determines the linewidth divergence as discussed above.

\subsection{Neutron diffraction }\label{SEC:Neutron}

\begin{figure}[bt!]
\includegraphics[clip,width=\columnwidth]{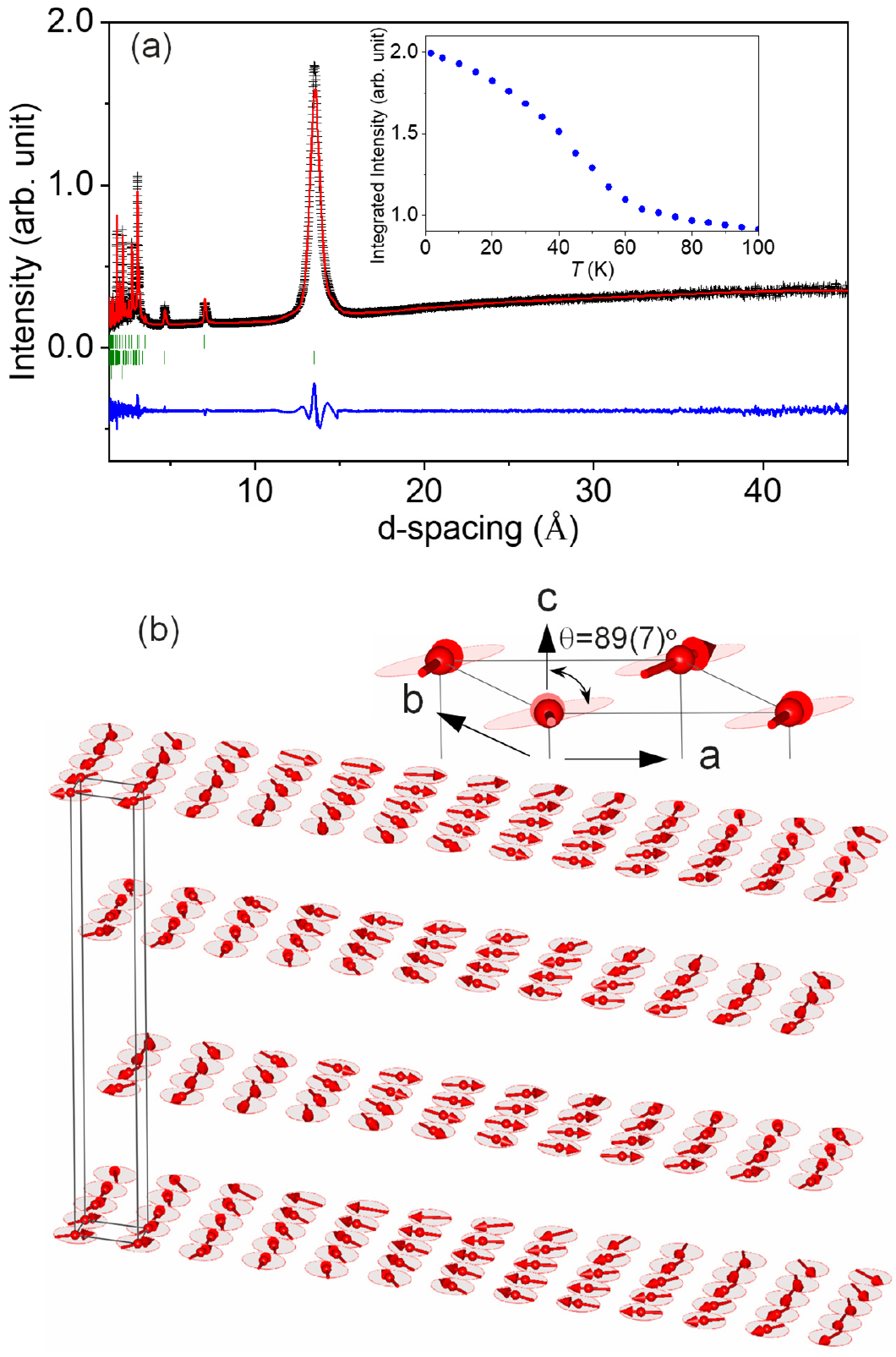}
\caption{(a) Rietveld refinement of the neutron diffraction data collected at $T=1.5$~K with the forward-scattering detector bank [$R_{\rm Bragg}({\rm nuclear})=6.24\%$ and $R_{\rm Bragg}({\rm magnetic})=3.48\%$]. The cross symbols and solid red line represent the experimental and calculated intensities, respectively, and the blue line below is the difference between them. Tick marks (green) indicate the positions of Bragg peaks: nuclear (top), magnetic (middle) and vanadium can (bottom). Inset shows integrated intensity of the diffraction pattern in the $d$-spacing range of $12-15$ ~\AA\, where the strongest magnetic satellite $(0.037,0.037,3/2)$ is observed, as a function of temperature. (b) Schematic representation of the long-period modulated magnetic structure of AgCrSe$_2$. }
	\label{Neutron}
\end{figure}

Neutron diffraction experiments were carried out on powdered polycrystalline material. The polycrystalline sample was characterized in full detail (see the Supplemental Material \cite{SM}). The susceptibility shows a maximum at about 50~K (frequently taken as the order temperature), a Weiss temperature around $70$~K and an effective moment of about $\mu_{\rm eff} = 3.52$~$\mu_{\rm B}$ in good agreement with earlier results on AgCrSe$_{2}$ powder \cite{Bongers,Gautam}. Refinement of the crystal structure was done in the non-centrosymmetric space group $R3m$. The obtained structural parameters are summarized in the Supplemental Material \cite{SM} in Table S1. The crucial step to achieve a good fitting quality was to refine anisotropic thermal parameters for Ag and Se2 (see Supplemental Material \cite{SM}). The large values of the $\beta_{11}$ and $\beta_{22}$ parameters likely indicate a presence of cation disorder over several positions on the tetrahedral holes as originally discussed by Engelsman \textit{et al.}\ in an early work \cite{Engelsman}. The non-centrosymmetric nature of the crystal structure of AgCrSe$_2$ implies that a Lifshitz-type invariant is allowed in the Landau free-energy decomposition similar to the well-known case of the room temperature multiferroic BiFeO$_3$ \cite{Kado}, which also has a polar trigonal structure. In the latter system, the presence of the Lifshitz invariant results in a long-period cycloidal magnetic ground state with the polar $c$ axis confined within the plane of spin rotations. At a microscopic level, this magnetic ground state is stabilized by antisymmetric exchange interactions associated with the polar structural distortions \cite{Zvezdin}. By analogy, one can assume that a similar situation might take place in AgCrSe$_2$. Indeed, a long-period cycloidal magnetic structure has been reported for this compound by Engelsman \textit{et al.}\ based on neutron diffraction data \cite{Engelsman}. However, the spins were found to be within the ($a$,$b$) - plane, which is not consistent with the mechanism where the cycloidal order is stabilized by the antisymmetric exchange. To verify this crucial point, we undertook additional neutron powder diffraction measurements, targeting to explore the possibility of the ground state similar to BiFeO$_3$. In agreement with \cite{Engelsman}, the magnetic reflections appearing below 50~K (Fig.~\ref{Neutron}(a), inset) can be indexed by the $k=(0.037,0.037,3/2)$ propagation vector. The reflections were found to be notably broader than the nuclear diffraction peaks indicating a finite correlation length of the magnetic order. This might be associated with the size of magnetic domains naturally expected for this propagation vector with three-arm star. The quantitative refinement of the magnetic structure [Fig.~\ref{Neutron}(a)] was done in a polar coordinate system where the spatial position of the cycloidal plane was controlled by the polar angle $\theta$ which was freely refined along with the size of Cr moments. The obtained magnetic structure [Fig.~\ref{Neutron}(b)] is very close to the one reported by Engelsman \textit{et al.}\ \cite{Engelsman} with no statistically significant deviation of the cycloidal plane from the ($a$,$b$) - plane, $\theta=89(7)^{\circ}$. The standard deviation for $\theta$ is however rather big indicating that a small deviation might exist but more precise single crystal data are required to confirm that. Thus the magnetic structure confirmed in the present study strongly indicates that the long-period modulated state in AgCrSe$_2$ is stabilized by competing symmetric exchange interactions rather than a competition between symmetric and antisymmetric exchange like in BiFeO$_3$ or in other systems with non-centrosymmetric crystal structures \cite{Pfleiderer,Takada,Hara}. The ordered moment at $T=1.5$~K is $2.53(5)~\mu_{\rm B}$, which is significantly smaller than the expected value for the $S=3/2$ Cr$^{3+}$ cations and the saturation value obtained in the high magnetic field magnetization measurements (Figs.~\ref{MperpxH} and \ref{MparaxH}). This experimental result might be related to a presence of a disordered component associated with the structural disorder or might indicate a presence of strong spin fluctuations persisting down to very low temperatures.

\begin{figure}[bt!]
\includegraphics[clip,width=\columnwidth]{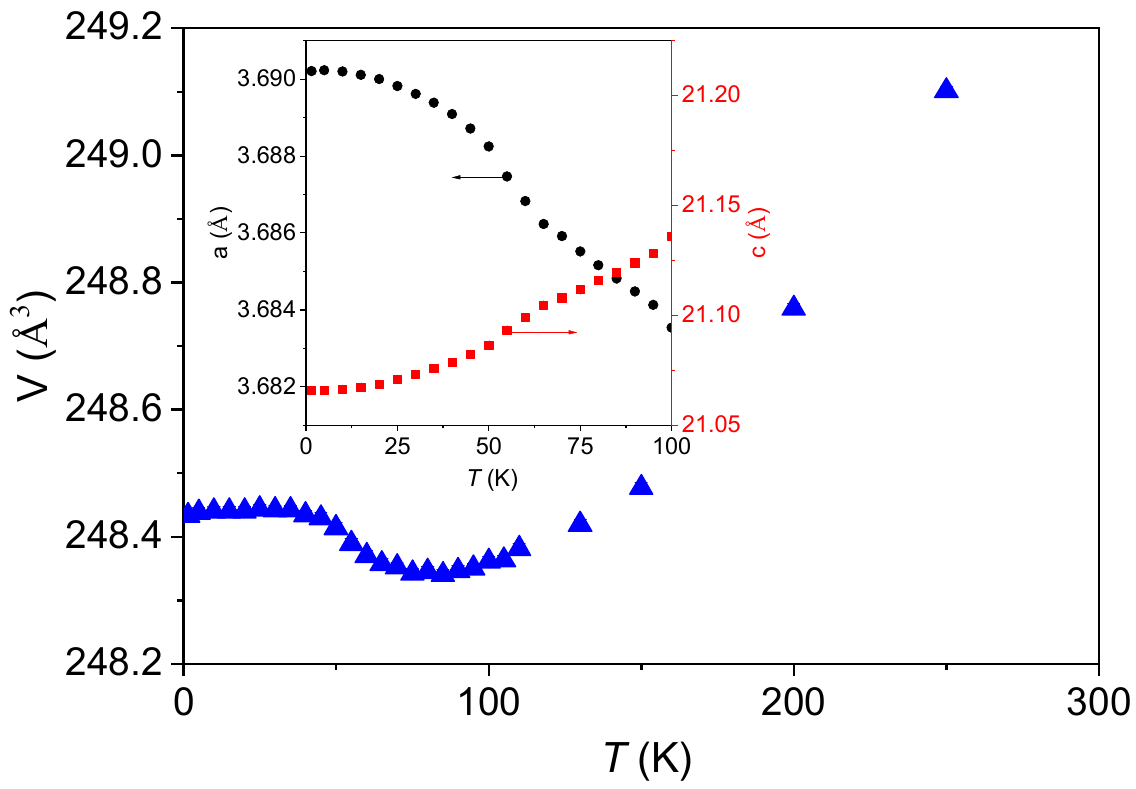}
\caption{Unit cell volume as a function of temperature. Inset shows the temperature dependence of the unit cell parameters.}
	\label{Neutron2}
\end{figure}

Above the transition temperature to the magnetically ordered state, a pronounced diffuse magnetic scattering is observed (see Supplemental Material \cite{SM}) revealing the presence of well-defined spin correlations. The diffuse scattering remains up to 100~K indicating that the spin correlations start to develop at temperatures significantly higher than the magnetic transition temperature. This is a typical behavior for magnetically frustrated systems which further supports the conclusion that the modulated state in AgCrSe$_2$ is stabilized by competing exchange interactions. The developing spin correlations and the transition to the ordered state are accompanied by a visible lattice effect (Fig.~\ref{Neutron2}) with a negative thermal expansion in the temperature range of $50~{\rm K} < T < 100$~K.  This evidences that lattice distortions play an essential role in releasing frustration and establishing the long-range ordered magnetic ground state.

\subsection{Competing Heisenberg exchange between Cr ions}\label{SEC:theory}

Delafossite compounds containing chromium inside a distorted chalcogenide octahedra typically cannot be described with an exchange model containing nearest-neighbor interactions only \cite{ushakov}. A finite saturation field for the magnetization $M(H)$ together with a positive Curie-Weiss temperature of the susceptibility $\chi(T)$ demonstrates that we are faced with at least two competing exchange parameters. Preliminary density-functional calculations \cite{Kim} suggest that the dominant exchange interactions are ferromagnetic $J_1<0$ between nearest neighbors and antiferromagnetic between third-nearest neighbors with $J_3={\cal O}(|J_1|)$. Together with a small but finite single-ion anisotropy $D>0$ we thus use a minimal Hamiltonian of the form
\begin{align}
	{\cal H}
	&=
	\sum_{i=1}^\nu{\cal H}(i),
	\label{eqn:cr:ham}
	\\
	{\cal H}(i)
	&=
	D\left(S_i^z\right)^2
	-g\mu_\text B\mu_0\vec H\vec S_i
	\nonumber\\
	&
	+\frac12J_1\sum_{j\in\text{nn}(i)}\vec S_i\vec S_j
	+\frac12J_3\sum_{j\in\text{3n}(i)}\vec S_i\vec S_j,
	\nonumber
\end{align}
for $\nu$ magnetic ions, ignoring possible symmetry-allowed symmetric and antisymmetric exchange anisotropies as well as second-neighbor and inter-plane exchange.

A high-temperature expansion for the magnetic susceptibility \cite{schmidt,johnston} yields
\begin{align}
	\theta_\text{CW}^\parallel
	&=
	-\frac{s(s+1)}{3k_\text B}z(J_1+J_3)
	+\frac{(2s-1)(2s+3)}{15k_\text B}D,
	\\
	\theta_\text{CW}^\perp
	&=
	-\frac{s(s+1)}{3k_\text B}z(J_1+J_3)
	-\frac{(2s-1)(2s+3)}{30k_\text B}D
\end{align}
with $s=3/2$ and $z=6$. Here, the symbols $\parallel$ and $\perp$ indicate the field direction relative to the direction of the single-ion anisotropy.

The saturation field $\mu_0H_\text{S}^\parallel$ of the magnetization $M(H)$ is the strength of the applied magnetic field where the sample reaches the fully polarized state with moment $M_\text{S}$ per Cr ion. It is given by \cite{schmidt1,schmidt2}
\begin{equation}
	\mu_0H_\text{S}^\parallel
	=
	\frac{2s^2}{M_\text{S}}\left(D+J(0)-J(\vec Q)\right),
	\label{eqn:cr:hsat}
\end{equation}
where $M_\text{S}=g_\parallel\mu_\text Bs$ and $\vec Q$ is the ordering vector of the magnetic phase in the $xy$ plane perpendicular to the field. Here we assume $\vec H$ points into the direction of the single-ion anisotropy defining the global $z$ direction. The Fourier transform of the exchange tensor in the $xy$ plane is given by
\begin{equation}
    J(\vec k)=\frac{1}{\nu}
    \sum_{\langle ij\rangle}J_{ij}
    e^{-{\rm i}\vec k(\vec R_{i}-\vec R_{j})}
    =\frac{1}{2}\sum_{n}J_{n}e^{-{\rm i}\vec k\vec R_{n}}
\end{equation}
where the sum runs over all bonds $n$ connecting a fixed site $i$ with its neighbors. Minimizing the classical energy $\nu s^2J(\vec Q)$ for the Hamiltonian~(\ref{eqn:cr:ham}) above with respect to the components of the ordering vector yields a series of magnetic phases. With $\alpha:=J_3/J_1$, $J_1<0$ we obtain the following two:

\paragraph{Ferromagnetic phase.}

For $\alpha\ge0$ the ground state trivially is a ferromagnet, i.\,e., $\vec Q=0$, and $J(0)=3(J_1+J_3)$. The saturation field in this case is determined by the single-ion anisotropy $D$ alone. This remains true also for $-1/4\le\alpha<0$: even a frustrating AFM third-neighbor exchange $J_3>0$ with $J_3<|J_1|/4$ doesn't destroy the fully polarized ground state.

\paragraph{Spiral phase.}

For $\alpha<-1/4$ or $J_3>|J_1|/4$ the classical ground state is a spiral along $[110]$ with an incommensurate $\vec Q$ vector,
\begin{align}
	\vec Q
	&=
	\left(
		2\cos^{-1}\left(\frac1{2\sqrt2}
		\sqrt{5-\frac{\sqrt{\alpha(\alpha-2)}}\alpha}
	\right),
	\right.
	\nonumber\\&\phantom=\left.
	\frac2{\sqrt3}\cos^{-1}\left(
	\frac{\sqrt{5-\frac{\sqrt{\alpha(\alpha-2)}}\alpha}}
	{2\sqrt2\left(\sqrt{\alpha(\alpha-2)}-\alpha\right)}
	\right)
	\right).
	\label{eqn:cr:q}
\end{align}
Correspondingly we obtain
\begin{equation}
	J(\vec Q)
	=
	-\frac{J_1}4\left(1+\frac{J_1}{2J_3}+5\frac{J_3}{J_1}\right)
	-\frac{J_3}4\left(1-\frac{2J_1}{J_3}\right)^{3/2}
\end{equation}
for the ground-state energy and
\begin{align}
	\mu_0H_\text{S}^\parallel
	&=
	\frac{2s^2}{M_\text{S}}\left[D+
	\frac{J_1}4\left(13+\frac{J_1}{2J_3}+17\frac{J_3}{J_1}\right)
	\right.\nonumber\\&\phantom{=\frac{2s^2}{M_\text{S}}}\left.
	+\frac{J_3}4\left(1-\frac{2J_1}{J_3}\right)^{3/2}
	\right]
	\label{eqn:cr:hsatinc}
\end{align}
for the saturation field. This is the case for AgCrSe$_2$.

Experimentally, we have obtained a Curie-Weiss temperature $\theta =+75{\rm ~K}$, approximately independent of the field direction. The saturation field directed parallel to the crystallographic $c$ direction is obtained as $\mu_0H_\text{S}^\parallel=25.1~{\rm T}$ with a saturation moment $M_\text{S}=2.8~\mu_\text B/\text{Cr ion}$. Ignoring the small single-ion anisotropy, justified by the directional independence of $\theta$, we obtain $J_1=-1.73~{\rm meV}=-20k_\text B{\rm K}$ and $J_3=0.86~{\rm meV}=10k_\text B{\rm K}$.

Two remarks are due on the size of the single-ion anisotropy $D$. Although we find that $D$ must be small compared to the absolute values of the exchange constants, it is nonzero. A finite $D>0$ along the $c$ direction ``forces'' the Cr moments into the ($a$,$b$) - plane. Because of that, a saturation field in the ($a$,$b$) - plane does not depend on $D$ \cite{abarzhi}, therefore we may obtain an estimate of $D$ via
\begin{equation}
    D
    =
    \frac{\mu_0}{2s^2}M_\text{S}\left(H_\text{S}^\parallel-H_\text{S}^\perp\right)
    =
    0.05\,{\rm meV}
    =
    0.6k_\text B{\rm K}
\end{equation}
with $\mu_0H_\text{S}^\perp=23.7{\rm ~T}$ in the ($a$,$b$) - plane.

The meta-magnetic transition we observe at a field $\mu_{0}H^{+}=5~\rm T$ at low temperatures supports this scenario: Any field having a component perpendicular to the direction of $D$ breaks the rotational symmetry of the Hamiltonian around $D$. For small fields, the ordered moments remain in the plane and form a ``distorted helix'' until $H^{+}$is reached: a transition to a fan phase occurs. For a purely AFM nearest-neighbor exchange $J$, the critical field is given by $\mu_0H^{+}=\left(2s^2/M_\text{S}\right)\sqrt{DJ}$ \cite{abarzhi,rastelli}, and we expect a similar scaling with an effective exchange constant $J_\text{eff}\gg D$ derived from the exchange structure of our model, Eq.~\ref{eqn:cr:ham}.

\subsection{\boldmath  Single crystal $H-T$ phase diagram}\label{SEC:Phase}

\begin{figure}[bt!]
\includegraphics[clip,width=\columnwidth]{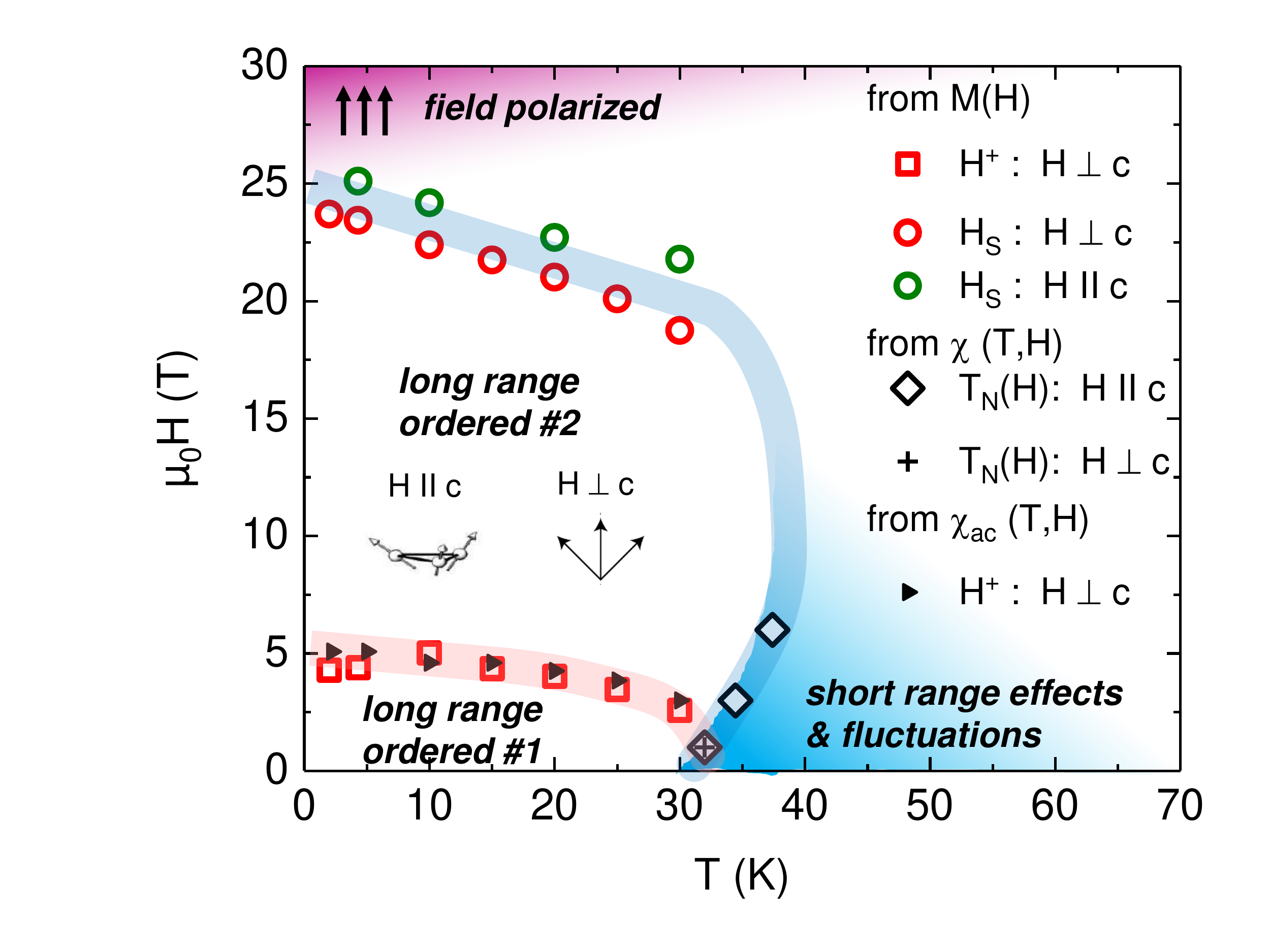}
\caption{$H-T$ phase diagram of AgCrSe$_{2}$, obtained from various methods for magnetic fields $H  \perp c$ and $H \parallel c$. The ordered state \#1 represents the cycloidal phase for $H  \perp c$ below $H^{+}(T)$ (see Fig.~\ref{Neutron}(b)), whereas phase \#2 represents a planar fan phase for in-plane fields (above $H^{+}(T)$) and a more cone-like phase for fields in $c$ direction. }
\label{PhaseDiagram}
\end{figure}

The magnetic exchange interactions in AgCrSe$_{2}$ are very different from those in the oxygen-based Cr-delafossites. Whereas the latter are characterized by a predominant and a very strong AFM nearest-neighbor interaction, AgCrSe$_{2}$ features an interplay of FM nearest neighbor and AFM third-nearest-neighbor interactions. As a consequence, AgCrSe$_{2}$ shows a cycloidal order with large modulation length (about 164 \AA ) along the (1,1,0) direction \cite{Engelsman}, whereas the oxygen-based Cr-delafossites exhibit a robust $120^{\circ}$ order on the triangular lattice. In AgCrSe$_{2}$, the application of magnetic fields first results in a modulation of the magnetic structure and, due to the smaller magnitude and different nature of the exchange interactions, it is possible to ferromagnetically align the moments in rather small fields of only about 25~T. That is different in the Cr-oxygen-delafossites, where the saturation fields are estimated to be an order of magnitude higher, e.g.\ 240 T in CuCrO$_{2}$ \cite{Reyes}. AgCrSe$_{2}$ is thus characterized by a favorable combination of competing interactions between Cr ions and thermal excitations, which finally leads to a tunable magnetic ground state and the emergence of magnetic fluctuations at the verge of magnetic order.

Figure \ref{PhaseDiagram} shows the $H-T$ phase diagram of AgCrSe$_{2}$ for both directions, $H  \perp c$ and $H \parallel c$. We note that the $H-T$ phase diagram of polycrystalline AgCrSe$_{2}$, shown in the Supplemental Material \cite{SM}, displays some marked differences to that of the single crystal due to the strong 2D nature of the exchange interactions.
In high magnetic fields, AgCrSe$_{2}$ displays an isotropic behavior, i.e.\ almost the same saturation fields are found for both field directions. For $H \perp c$ a meta-magnetic phase transition appears at about $H^{+}=5$~T, which corresponds to about $1/5$ of the saturation field. For spin $1/2$ triangular lattices with $120^{\circ}$ order a metamagnetic transition at a value of $1/3$ is predicted \cite{Kawamura,Seabra} and found for example in Ba$_{3}$CoSb$_{2}$O$_{9}$ \cite{Shirata}. Above $H^{+}$ our results suggest a planar fan-type magnetic order in AgCrSe$_{2}$. For fields in $c$ direction our data are consistent with a continuous tilting of the moments out of the plane where the tilt angle increases upon increasing the field. These results motivate future neutron diffraction experiments on single crystals to solve the magnetic structure in applied magnetic fields.

Moreover, at high temperatures we have revealed a region of competing interactions characterized by magnetic fluctuations before AgCrSe$_{2}$ orders magnetically at lower temperatures. These magnetic fluctuations also influence the electronic transport properties of AgCrSe$_{2}$ \cite{Zhang}. Inelastic neutron scattering studies are needed to investigate the excitation spectrum of these fluctuations.

\section{SUMMARY AND CONCLUDING REMARKS}

In summary, we studied the $H-T$ phase diagram of AgCrSe$_{2}$ in great detail. We identified paramagnetic, long ranged ordered and field polarized regimes.
Whereas the Cr oxygen delafossites crystallize in the $R\bar3m$ space group and are characterized by a very strong AFM nearest-neighbor interaction, which then usually leads to a planar $120^{\circ}$ order, the Cr-sulfur and Cr-selenide delafossites display transitions from the $R\bar3m$ group to lower symmetries. Furthermore, higher order magnetic exchange interactions play an important role, which finally leads to a larger variation in the magnetic ground states; $120^{\circ}$, stripe, spin-spiral, and cycloidal-type ordered states are observed. In addition to the symmetry reduction from $R\bar3m$ to $R3m$ at high temperatures, a further symmetry reduction is usually found for the Cr sulfur delafossites at the onset of magnetic order. This effect is not observed in AgCrSe$_{2}$. In this respect, AgCrSe$_{2}$ is a model system for the study of a planar Cr triangular lattice with dissimilar Heisenberg-like interactions.

Our comprehensive study with local and bulk probes shows that AgCrSe$_{2}$ is an anisotropic cycloidal magnet with emerging anisotropic laterally extended magnetic fluctuations. For magnetic fields applied in the ($a$,$b$) - plane, we find a metamagnetic phase transition, which is not present for fields parallel $c$. Wheras we could study the magnetic structure in zero-field by neutrons diffraction, the magnetic structure in magnetic fields is still not known. From our data we also cannot rule out chiral magnetic textures in applied magnetic fields, which is certainly an attractive project left for future neutron studies.

The experimentally established $H-T$ phase diagram is consistent with a planar Heisenberg model, in which the nearest-neighbor interaction, unlike in the Cr-oxygen-delafossites, is ferromagnetic and the third-nearest-neighbor interaction is antiferromagnetic. In AgCrSe$_{2}$ the interactions among the Cr ions have just the right strength to be tuned by experimentally accessible magnetic fields. Based on our theoretical model and on the temperature dependence of the lattice parameters obtained from neutron diffraction, we suspect that the magnetic properties might be easily tuned by hydrostatic pressure or uniaxial stress. In this respect, AgCrSe$_{2}$ is a model system for the study of a planar Cr$^{3+}$ triangular lattice with dissimilar Heisenberg-like interactions.

\begin{acknowledgments}

We acknowledge support from the Deutsche Forschungsgemeinschaft (DFG) through the SFB 1143 and the W\"{u}rzburg-Dresden Cluster of Excellence on Complexity and Topology in Quantum Matter--$ct.qmat$ (EXC 2147, Project No.\ 390858490), as well as the support of the HLD at HZDR, a member of the European Magnetic Field Laboratory (EMFL). We gratefully acknowledge support from the European Research Council (through the QUESTDO project, 714193), the Leverhulme Trust, and the Royal Society. We thank the Elettra synchrotron for access to the APE-HE beamline under proposal number 20195300. The research leading to this result has been supported by the project CALIPSOplus under Grant Agreement 730872 from the EU Framework Programme for Research and Innovation HORIZON 2020. Part of this work has been performed in the framework of the Nanoscience Foundry and Fine Analysis (NFFA-MUR Italy Progetti Internazionali) project (www.trieste.NFFA.eu). We thank F.\ Mazzola for useful discussions and U.\ Nitzsche for technical support.  We thank O.\ Stockert, C.\ Geibel and A.~P.\ Mackenzie for fruitful discussions. We thank C.\ Klausnitzer, H.\ Rave and R.\ Hempel-Weber for technical support for magnetization measurements.

\end{acknowledgments}

\bibliography{AgCrSe2_VER30}

\end{document}